\documentclass[aps,prl,amsmath,amssymb,superscriptaddress,lengthcheck]{revtex4-1}

\newcommand{\ncco}{$\textrm{Nd}_{2-x}\textrm{Ce}_x\textrm{Cu}\textrm{O}_{4}$}

\newcommand{\bscco}{$\textrm{Bi}_2\textrm{Sr}_2\textrm{Ca}\textrm{Cu}_2\textrm{O}_{8+\delta}$}
\newcommand{\bsco}{$\textrm{Bi}_2\textrm{Sr}_{2-x}\textrm{La}_x\textrm{Cu}\textrm{O}_{6+\delta}$}
\newcommand{\ybco}{$\textrm{Y}\textrm{Ba}_2\textrm{Cu}_3\textrm{O}_{6+\delta}$}

\newcommand{\lnbsco}{$\textrm{(La,Nd)}_{2-x}\textrm{(Ba,Sr)}_x \textrm{Cu}\textrm{O}_{4}$}

\newcommand{\hg}{$\textrm{Hg}\textrm{Ba}_2\textrm{Cu}\textrm{O}_{4+\delta}$}
\newcommand{\qco}{$Q_{CO}$}
\usepackage{graphicx}
\usepackage{dcolumn}
\usepackage{bm}
\usepackage{relsize}
\usepackage{xcolor,color}
\begin{document}


\title{Coupling between dynamic magnetic and charge-order correlations in the cuprate superconductor \ncco}

\author{E.\,H.\,da Silva Neto}
\email[]{ehda@ucdavis.edu}
\affiliation{Department of Physics, University of California, Davis, California 95616, USA}
\affiliation{Quantum Matter Institute, University of British Columbia, Vancouver, British Columbia V6T 1Z4, Canada}
\affiliation{Max Planck Institute for Solid State Research, Heisenbergstrasse 1, D-70569 Stuttgart, Germany}

\author{M.\,Minola}
\affiliation{Max Planck Institute for Solid State Research, Heisenbergstrasse 1, D-70569 Stuttgart, Germany}

\author{B.\,Yu}
\affiliation{School of Physics and Astronomy, University of Minnesota, Minneapolis, Minnesota 55455, USA}

\author{W.\,Tabis}
\affiliation{Laboratoire National des Champs Magn\'etiques Intenses, CNRS UPR-3228, 31400, Toulouse, France}
\affiliation{AGH University of Science and Technology, Faculty of Physics and Applied Computer Science, 30-059 Krakow, Poland}
\affiliation{Institute of Solid State Physics, TU Wien, 1040 Vienna, Austria}

\author{M.\,Bluschke}
\affiliation{Max Planck Institute for Solid State Research, Heisenbergstrasse 1, D-70569 Stuttgart, Germany}
\affiliation{\mbox{Helmholtz-Zentrum Berlin f\"{u}r Materialien und Energie, BESSY II, Albert-Einstein-Str. 15, 12489 Berlin, Germany}}

\author{D.\,Unruh}
\affiliation{Department of Physics, University of California, Davis, California 95616, USA}

\author{H.\,Suzuki}
\affiliation{Max Planck Institute for Solid State Research, Heisenbergstrasse 1, D-70569 Stuttgart, Germany}

\author{Y.\,Li}
\affiliation{School of Physics and Astronomy, University of Minnesota, Minneapolis, Minnesota 55455, USA}

\author{G.\,Yu}
\affiliation{School of Physics and Astronomy, University of Minnesota, Minneapolis, Minnesota 55455, USA}

\author{D.\,Betto}
\affiliation{European Synchrotron Radiation Facility, 71 Avenue de Martyrs, CS40220, F-38043 Grenoble Cedex 9, France}

\author{K.\,Kummer}
\affiliation{European Synchrotron Radiation Facility, 71 Avenue de Martyrs, CS40220, F-38043 Grenoble Cedex 9, France}

\author{F.\,Yakhou}
\affiliation{European Synchrotron Radiation Facility, 71 Avenue de Martyrs, CS40220, F-38043 Grenoble Cedex 9, France}

\author{N.\,B.\,Brookes}
\affiliation{European Synchrotron Radiation Facility, 71 Avenue de Martyrs, CS40220, F-38043 Grenoble Cedex 9, France}

\author{M.\,Le\,Tacon}
\affiliation{Max Planck Institute for Solid State Research, Heisenbergstrasse 1, D-70569 Stuttgart, Germany}
\affiliation{Institut f\"ur Festk\"{o}rperphysik, Karlsruher Institut f\"ur Technologie, 76201 Karlsruhe, Germany}

\author{M.\,Greven}
\affiliation{School of Physics and Astronomy, University of Minnesota, Minneapolis, Minnesota 55455, USA}

\author{B.\,Keimer}
\affiliation{Max Planck Institute for Solid State Research, Heisenbergstrasse 1, D-70569 Stuttgart, Germany}

\author{A.\,Damascelli}
\email[]{damascelli@physics.ubc.ca}
\affiliation{Quantum Matter Institute, University of British Columbia, Vancouver, British Columbia V6T 1Z4, 
Canada}
\affiliation{Department of Physics \& Astronomy, University of British Columbia, Vancouver British Columbia, V6T 1Z1, Canada}

\begin{abstract}

Charge order has now been observed in several cuprate high-temperature superconductors. We report a resonant inelastic x-ray scattering experiment on the electron-doped cuprate \ncco~that demonstrates the existence of dynamic correlations at the charge order wave vector. Upon cooling we observe a softening in the electronic response, which has been predicted to occur for a $d$-wave charge order in electron-doped cuprates. At low temperatures, the energy range of these excitations coincides with that of the dispersive magnetic modes known as paramagnons. Furthermore, measurements where the polarization of the scattered photon is resolved indicate that the dynamic response at the charge order wave vector primarily involves spin-flip excitations. Overall, our findings indicate a coupling between dynamic magnetic and charge-order correlations in the cuprates.


\end{abstract}

\maketitle

In addition to the long-studied superconducting (SC), antiferromagnetic (AF), and pseudogap phases, the copper-based high-temperature superconductors (cuprates) also feature charge order (CO) correlations. The CO is a periodic organization of low-energy electronic states and it is ubiquitous to all cuprate families \cite{tranquada_evidence_1995, Hoffman_2002, Howald_2003, Abbamonte_2005, Wu_2011, Ghiringhelli_CDW_2012, Chang_CDW_2012, Achkar_2012, Comin_CDW_2014, daSilvaNeto_CDW_2014, Hashimoto_2014, Tabis_2014, daSilvaNeto_Science_2015, daSilvaNeto_SciAdv_2016, Tabis_PRB, PRX_NCCO}. 
Early theoretical works that predicted an instability toward an intertwined pattern of charge and spin order, known as stripes \cite{zaanen_charged_1989, Kivelson1998}, were first confirmed by neutron scattering experiments in the La-based family, \textit{i.e.}\,\lnbsco~\cite{tranquada_evidence_1995}. In more recent years, the observation of CO in 
other cuprate families has led to new theories that suggest a tight link between the emergence of CO and AF fluctuations \cite{Davis_DHLee_2013, Efetov_nat_phys_2013,Sachdev_PRL_2013, Wang_Chubukov_PRB_2014}. 
However, the possibility of such an interplay in non-stripe materials remains controversial. 
Several studies show no clear correlation between the doping evolution of the CO with that of the AF properties \cite{Santi_2014, Hucker_PhysRevB.90.054514, daSilvaNeto_SciAdv_2016}, except for the case of Zn-doped \ybco~(YBCO), where magnetic order is nucleated by Zn impurities at the expense of the CO \cite{Santi_PRL2013}. Still, a clear connection between CO and AF correlations remains elusive due to the lack of experiments that simultaneously resolve the excitations of both charge and spin degrees of freedom. The electron-doped cuprate \ncco~(NCCO), featuring more prominent AF correlations than its hole-doped counterparts \cite{Motoyama_Spin_2007}, also develops CO \cite{daSilvaNeto_Science_2015}, making it an ideal system where to investigate the relation between AF and CO correlations. 

Although static CO has been detected by a variety of experiments that probe electrons at long time scales, momentum-resolved evidence for dynamic CO correlations has proven to be more elusive. In YBCO, where the CO is the strongest (apart from the special case of stripes in La-cuprates), Cu-$L_3$ resonant inelastic x-ray scattering (RIXS) shows that the CO is quasi-elastic within $130$\,meV \cite{Ghiringhelli_CDW_2012}, and non-resonant inelastic X-ray scattering indicates that the lattice distortion associated with the CO is static within $1.4$\,meV \cite{LeTacon2014}. 
However, several experiments \cite{Wu_2011,Wu_2015,Gerber_2015, Chang2016, Jang2016} showed that the intensity and correlation length of the CO, in a narrow doping range of YBCO, are dramatically enhanced in magnetic fields above $12$\,T, suggesting that the short-range ($\approx 65$\,\AA) CO at zero field is likely a precursor state to the high field CO.
Even shorter-range CO correlations ($\approx 25$\,\AA) are observed in most other cuprates, including electron-doped materials \cite{Comin_CDW_2014, daSilvaNeto_CDW_2014, Tabis_2014, daSilvaNeto_Science_2015, daSilvaNeto_SciAdv_2016, Miao_2017, PRX_NCCO}. Thus, it is possible that the zero-field CO correlations in the cuprates are primarily dynamic in nature \cite{Kivelson_RMP_2002, Caplan_PRB_2015, Wu_PRB_2016} and their observation by static probes is the result of disorder pinning. More recently, a RIXS study reported the observation of CO excitations ($\approx 50$\,meV) in \bscco~\cite{Chaix2017}, although a coupling to dynamic 
magnetic correlations was not reported.

Over the last few years, Cu-$L_3$ resonant X-ray scattering, either in energy-integrated mode (EI-RXS) or in energy-resolved inelastic mode (RIXS), has become the tool of choice for the detection of CO in the cuprates. In both cases, tuning of the photon energy to the Cu-$L_3$ edge enhances the sensitivity of the scattering cross-section to the low-energy electronic states that derive from the CuO$_2$ planes. In a typical EI-RXS measurement,
all photons scattered into a reciprocal-lattice element are picked up by the
detector, whereas in RIXS the scattered photons are analyzed by a spectrometer that resolves their energy. 
Consequently, the EI-RXS measurements (e.g.\,\cite{Comin_CDW_2014, daSilvaNeto_CDW_2014, Santi_2014, Thampy_2014, daSilvaNeto_Science_2015}) cannot rule out the contribution of inelastic scattering to the broad CO peaks in momentum space.
Here, we exploit the high-resolution RIXS instrument at the European Synchrotron Radiation Facility to uncover the presence of dynamic correlations at the charge order wave vector (\qco) in both non-SC ($x=0.106$) and SC ($x=0.145$) NCCO, using the same samples studied in prior EI-RXS work \cite{daSilvaNeto_SciAdv_2016}.
We find that a large contribution to the dynamic correlations at \qco~occurs in the same energy range spanned by the magnetic excitations.
By resolving the polarization of the scattered photons, we find that this enhancement of the dynamic response at \qco~is mostly due to spin-flip processes, thus showing a direct coupling between dynamic magnetic and charge-order correlations in NCCO. 

\begin{figure}
\includegraphics[width=85mm]{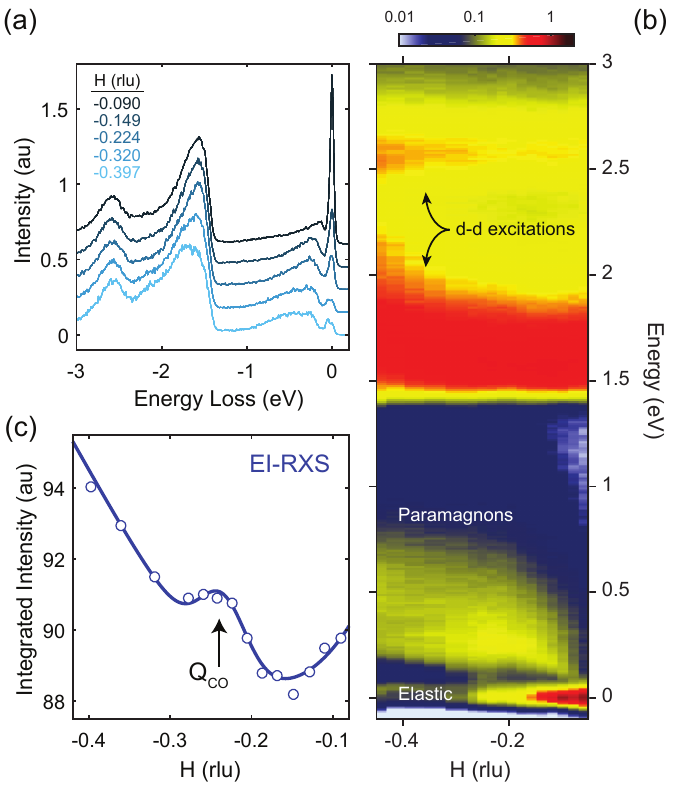}
\caption{\label{fig:1} 
(a), Cu-$L_3$ RIXS spectra for selected values of in-plane momentum transfer H. The curves are vertically offset for clarity. (b), Energy-momentum colormap of RIXS excitations. (c), EI-RXS momentum dependence obtained from the integrated RIXS spectra in the $(-0.06,10)$\,eV range. The measurements were performed on the non-SC sample, at $25$\,K, in $\sigma$ scattering geometry. The color scale in (b) is logarithmic and in the same units as (a).  
}
\end{figure}

Figure \ref{fig:1}(a) shows select Cu-$L_3$ RIXS spectra measured at different values of $H$, the in-plane momentum transfer in reciprocal lattice units (rlu), with an energy resolution of approximately $60$\,meV (full width at half maximum) \cite{SM}. This information can be compiled in a single color plot, Fig.\,\ref{fig:1}(b), which shows the energy-momentum structure of excitations in NCCO, including the elastic line ($E=0$\,eV) and dispersive excitations in the mid-infrared region (MIR, $100$-$500$\,meV), as well as $d$-$d$~($E > 1.3$\,eV, \cite{MMS_dd}) excitations.
To illustrate the relation between EI-RXS and RIXS, we integrate the RIXS spectra over a large energy range, $(-0.06,10)$\,eV. The result is a single momentum distribution curve, Fig.\,\ref{fig:1}(c), that emulates previous EI-RXS measurements of NCCO \cite{daSilvaNeto_SciAdv_2016}. Note that the CO peak constitutes only a small fraction of the integrated intensity ($\approx 1\%$, similar to actual EI-RXS measurements \cite{Comin_CDW_2014, daSilvaNeto_CDW_2014, Santi_2014, Thampy_2014, daSilvaNeto_Science_2015}) and that the large background comes from all other elastic and inelastic scattering within the energy-integration range.
Therefore, the peak at \qco~in an EI-RXS experiment is not necessarily restricted to elastic contributions, and it may originate from excitations with 
energies anywhere within a window of several electron-volts. Our RIXS experiments aim to dissect the inelastic spectrum of NCCO near \qco.


\begin{figure}
\includegraphics[width=83mm]{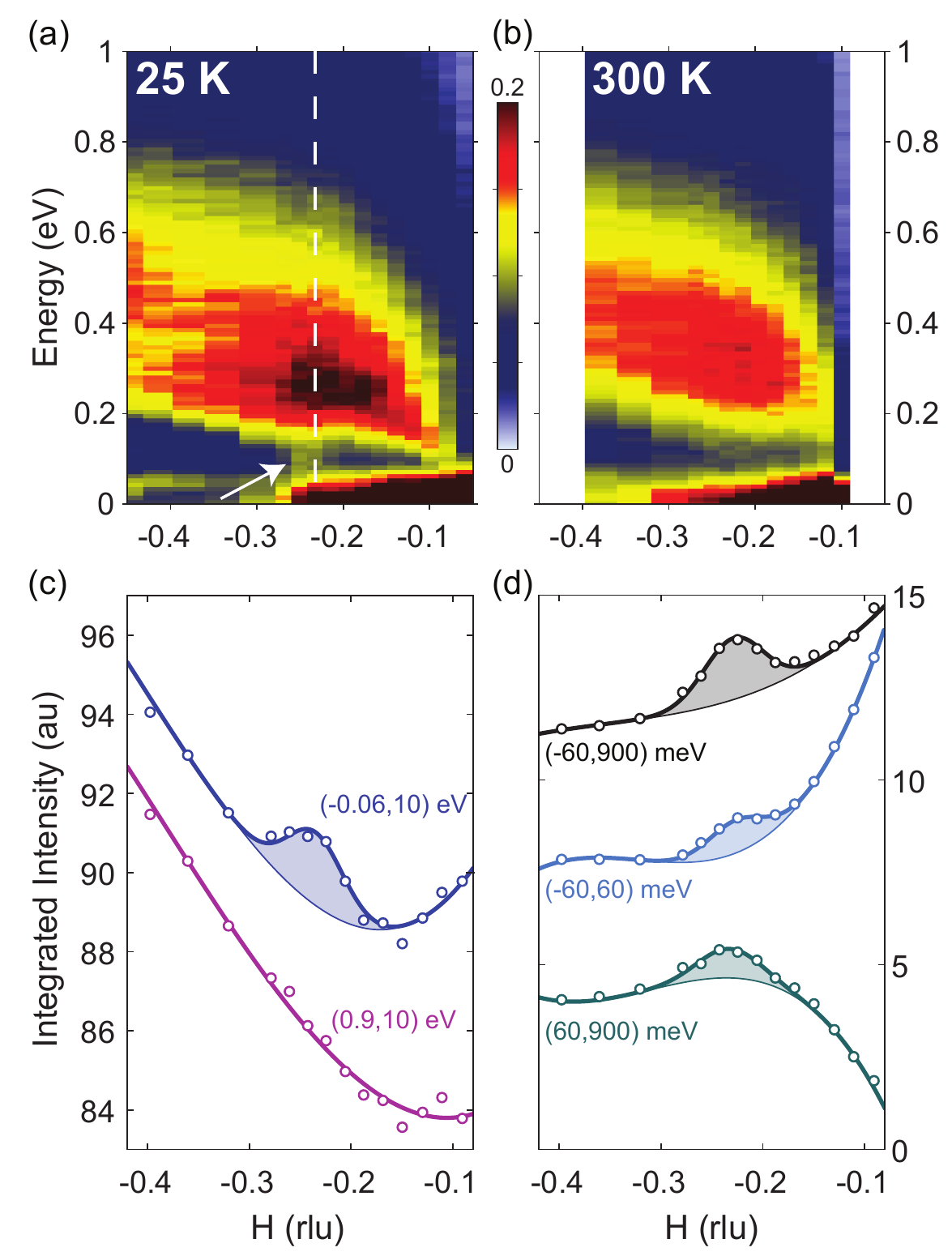}
\caption{\label{fig:2} 
Energy-momentum structure of the excitations in NCCO $0.106$ measured with $\sigma$ scattering at (a) $25$\,K and (b) $300$\,K. 
(c)-(d), $25$\,K RIXS signal integrated over different energy ranges showing a significant inelastic contribution to the peak at \qco~(black, top curve in (c), which is the sum of the lower two curves up).
The dashed line in (a) marks \qco~obtained from the energy-integrated data in (c) \cite{SM}. 
The variable pixel size in (a)-(b) reflects the values of H and E for which the raw data was acquired.
In (c)-(d) the data are open circles, while the thick lines are fits to the data to a polynomial function (thin lines) for the background plus a Gaussian for the peak, except for the magenta curve in (c), which is a polynomial fit. 
In (c)-(d) the data are vertically offset for clarity. 
}
\end{figure}

\begin{figure}
\includegraphics[width=83mm]{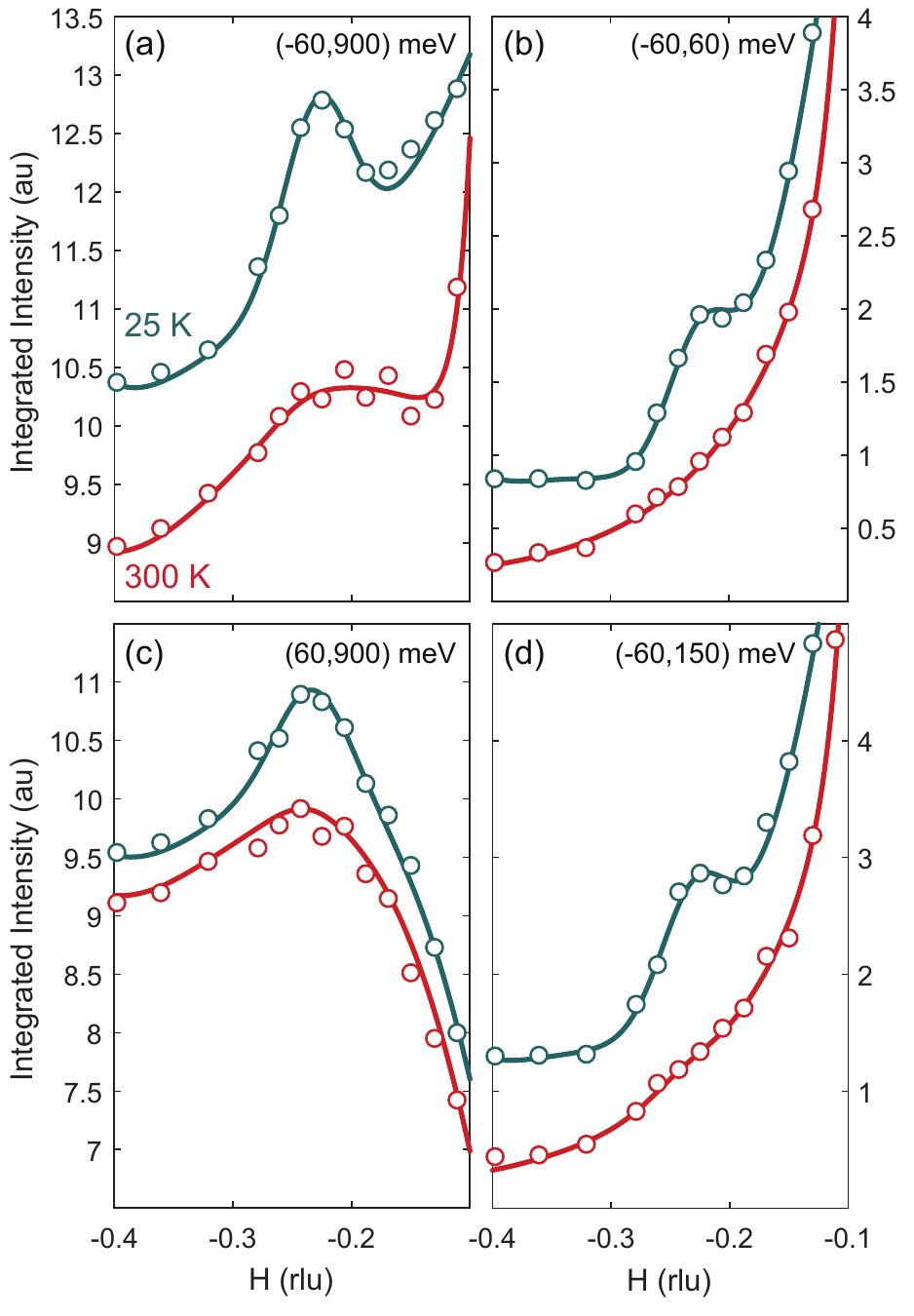}
\caption{\label{fig:3} 
(a)-(d), Energy-integrated RIXS for $25$\,K and $300$\,K in the energy ranges marked on the figure. All data on the non-SC sample in $\sigma$-geometry. The data are represented by open circles, while the lines represent a fit to data using a polynomial for the background plus a Gaussian function for the peak, except for the red curve in (b) which is simply a polynomial fit. 
The high temperature data have been vertically offset for clarity.
}
\end{figure}

Figure \ref{fig:2}(a) shows the detailed structure of the excitations in the non-SC NCCO sample at $25$\,K. 
Following previous EI-RXS measurements, we maximize the sensitivity to the CO by setting the polarization of the incoming photons parallel to the CuO$_2$ planes ($\sigma$ incoming polarization) in back scattering (using grazing incidence, $H<0$ in our convention) \cite{SM}.
With this geometry, broad dispersive modes in the MIR include contributions from charge, spin, and phonon excitations 
\cite{Ament_RevModPhys.83.705,Lucio_PhysRevLett.104.077002,LeTacon2011,Lee2014,YY_Peng_2015,Jia_PhysRevX.6.021020,Devereaux_PhysRevX.6.041019}. In particular, single spin-flip excitations in doped cuprates, have been named paramagnons, as they are the analogue to the magnon excitations in the AF undoped compound. Paramagnons have been detected in a variety of RIXS measurements and appear as damped but well-defined dispersive modes in the MIR region of the spectrum \cite{Hill_PhysRevLett.100.097001, Lucio_PhysRevLett.104.077002, LeTacon2011, Ishii_Soin_2014, Lee2014}. For NCCO, MIR charge modes that exist only near the zone center ($|H|<0.1$\,rlu) have also been reported \cite{Ishii_Soin_2014, Lee2014} but they are outside the scope of our study. We note that our measurements show an excellent agreement with the paramagnon dispersion obtained from previous RIXS measurements \cite{Lee2014, SM}. 
In our measurements, we find clear signals at \qco~near the paramagnon energies. This is manifest in the raw data as a dynamic signal at \qco~below the paramagnon energies and above the quasielastic line, marked by the white arrow in Fig.\,\ref{fig:2}(a) ($E\approx 100$\,meV),
accompanied by a scattering enhancement at higher energies ($E\approx 250$\,meV). Comparing this low-temperature measurement to its counterpart at $300$\,K, Fig.\,\ref{fig:2}(b), we find that these two features are mostly suppressed, in agreement with the temperature dependence of the CO obtained from the EI-RXS measurements \cite{daSilvaNeto_SciAdv_2016}. 
We also found a similar energy-momentum structure in measurements for $H > 0$, and it is confirmed on the SC NCCO sample near optimal doping ($T_c = 19$\,K) \cite{SM}. 


Not immediately clear from the data in Fig.\,\ref{fig:2}(a) is the presence of a CO peak in the quasi-elastic line. In order to clearly establish its presence and to quantify its strength relative to the inelastic signal at \qco, we separate the RIXS excitation spectrum into different energy regions by constructing energy-integrated momentum distribution curves. First, Fig.\,\ref{fig:2}(c) shows a comparison of the signal integrated over all measurable energies, $(-0.06,10)$\,eV range, to the curve obtained by integrating over the $(0.9,10)$\,eV range, which indicates that the peak at \qco~is fully contained below $0.9$\,eV.
Keeping in mind the energy resolution of these measurements ($\Delta E \approx 60$\,meV), in Fig.\,\ref{fig:2}(d) we decompose the \qco~signal into quasi-elastic, $(-60,60)$\,meV, and inelastic, $(60, 900)$\,meV, contributions. This analysis shows that roughly half of the peak observed in the EI-RXS measurements at low temperatures comes from inelastic scattering. Also note that this inelastic signal in the $60$ to $900$\,meV range seems to smoothly evolve from the quasi-elastic region (dashed line in Fig.\,\ref{fig:2}(a)). 

The decomposition in Fig.\,\ref{fig:2}(d) suggests the coexistence of both dynamic and static CO at low temperatures and the data in Fig.\,\ref{fig:3}(a) show that although suppressed, the short-range correlations at \qco~are still present at $300$\,K, in agreement with previous EI-RXS measurements \cite{daSilvaNeto_SciAdv_2016}. 
Remarkably, this suppression occurs unevenly over the energy spectrum. While at room temperature the quasi-elastic component at \qco~is completely absent, Fig.\,\ref{fig:3}(b), a small inelastic contribution remains in the $(60,900)$\,meV range, Fig.\,\ref{fig:3}(c). 
We further note that at $300$\,K there are no CO correlations in the $(-60,150)$\,meV range, Fig.\,\ref{fig:3}(d). This indicates that (i) the high-temperature inelastic \qco~signal must originate from the $(150,900)$\,meV energy range \cite{SM} where the paramagnons also exist, and (ii) that any temperature dependence below $150$\,meV is strictly due to electronic degrees of freedom since all significant known phonon modes lie below that energy. \cite{dAstuto_2002}.
Despite the relatively high Cu-$L_3$ RIXS energy-resolution of our measurements, we are not able to exclude the possibility that the quasi-elastic signal at \qco~is purely dynamic even at low temperatures. Nevertheless, the temperature dependence observed in our measurements is consistent with the scenario where high-temperature dynamic correlations at \qco~develop into static charge order at low-temperatures.

At first, it might seem difficult to state whether the $300$\,K MIR feature is a signature of the CO, or whether it appears in the energy-integrated curves 
due to the shape and details of the paramagnon dispersion. 
Nevertheless, we highlight that in NCCO the paramagnon dispersion shape remains substantially unchanged between $x=0.04$ and $x=0.145$ \cite{Lee2014}, so that the appearance of the MIR feature at the distinct \qco~of two samples with different doping levels strongly suggests a connection to CO (see \cite{SM} for data on SC NCCO).
In this context, we make three phenomenological observations: (i) the MIR enhancement is observed at \qco~in all possible scattering geometries (positive and negative $H$, and $\sigma$/$\pi$ scattering; see Fig.\,\ref{fig:2} and \cite{SM}), (ii) it follows \qco~as a function of doping \cite{SM}, and (iii) it could explain previous EI-RXS measurements that observe a residual peak above $300$\,K centered at the low-temperature \qco~\cite{daSilvaNeto_SciAdv_2016}. Additionally, a similar enhancement has been observed in Cu-$L_3$ RIXS measurements of \bsco~\cite{YY_Peng_2015}. Altogether these observations indicate that the MIR spectral enhancement at \qco~might be a robust feature in the cuprates.

Having established the basic phenomenology of the excitations at \qco, we further investigate their microscopic origin. At first glance, the anomalies of the electronic response near \qco~shown in Fig.\,\ref{fig:2}(a) seem to resemble the case of a conventional charge density wave with electron-phonon coupling. However, in that classic case, the softening of phonon modes driving the static order results in a transfer of spectral weight from high to low energies. Although a longitudinal optical-phonon anomaly occurs in SC NCCO at about \qco~\cite{dAstuto_2002}, there are no reports of a classic phonon softening. The data in Figs.\,\ref{fig:2} and \ref{fig:3} show, instead, a build-up of dynamic electronic correlations centered at \qco~with decreasing temperature. Therefore, it is possible that two related effects are at play: (i) a softening of the electronic response below $150$\,meV, (ii) concomitant with a spectral weight enhancement at higher energies, centered at $250$\,meV. Below we describe measurements that investigate the origin of (ii), whereas in regards to (i), we note that a softening of the charge susceptibility is predicted for a $d$-wave CO in electron-doped cuprates~\cite{Greco_PRB}.

\begin{figure}
\includegraphics[width=84mm]{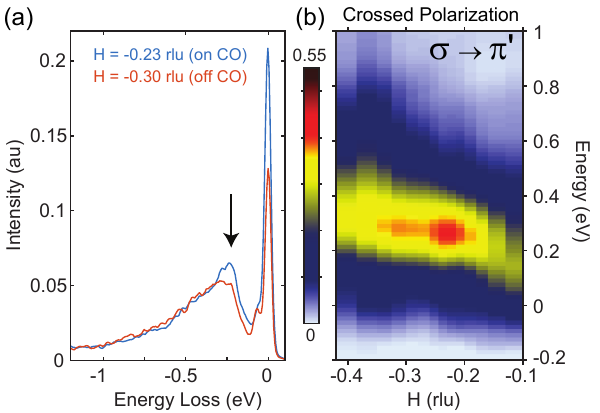}
\caption{\label{fig:4} (a) $25$\,K high energy-resolution ($35$\,meV) spectra taken on the non-SC NCCO, with incoming $\sigma$ polarization, for two H values, on and off \qco. (b) Energy-momentum structure for the excitations in the non-SC NCCO in the $\sigma\pi'$ channel, which is primarily composed of single spin-flip processes. The color plot in (b) was generated from eight polarization-resolved spectra~\cite{SM}.
}
\end{figure}

Now focusing on the MIR enhancement, it is imperative to determine whether 
it is purely due to charge scattering that coexists with the magnetic excitations, or whether it arises from additional spin-flip scattering at \qco. To investigate this issue, we first acquired high resolution RIXS spectra ($\Delta E \approx 35$\,meV), on and off \qco, Fig.\,\ref{fig:4}(a), which clearly corroborate our previous observation that the MIR CO signal exists at the same energy as the paramagnons. Although intriguing, this observation alone does not demonstrate the magnetic nature of the dynamic correlations at \qco. However, with the ability to resolve $\sigma\pi'$ scattering, where the prime is added to represent the polarization of the scattered photons, it is possible to isolate single spin-flip excitations. Fortunately, a newly developed polarimeter (based on the concept described in Ref.\,\cite{LB_RSI}) allowed us to perform such measurements, albeit with the compromise of a lower energy resolution ($\Delta E \approx 90$\,meV). Figure \ref{fig:4}(b) shows the RIXS $\sigma\pi'$ cross-section, which primarily follows the paramagnon dispersion, as expected for spin-flip scattering. Remarkably, the intensity of the paramagnons shows a measurable enhancement exactly at \qco, while no vestige of the dynamic signal was detected in the $\sigma\sigma'$ channel \cite{SM}. 

The combination of ultra-high resolution RIXS spectra and polarimetric decomposition indicates that the majority of the dynamic response at \qco~is only possible if it involves a flip of the electronic spin.
This last observation does not rule out the presence of dynamic charge correlations at energies lower than that of the paramagnons, as predicted in Ref.\,\cite{Greco_PRB}. Indeed, since that signal is much weaker than the MIR enhancement, Fig.\,\ref{fig:2}(a), its detection in the $\sigma\sigma'$ channel may still be beyond the sensitivity of the current state-of-the-art polarimetric RIXS instrumentation.
Note, however, that the fluctuations of a CO pattern will necessarily require a transfer of charge between neighboring sites, regardless of the mode of fluctuation -- phase shifts or amplitude enhancements. In fact, to conform to the underlying antiferromagnetic correlations, which are strong in electron-doped cuprates, these CO-related inter-site processes must involve a change of the spin degree of freedom. Since in this scenario a fluctuation of the charge-order pattern is coupled to a spin-flip process, it naturally follows that its excitation energy will largely be determined by the paramagnon energy scale at \qco. 
Nevertheless, lower-energy charge excitations near \qco~should remain possible since doping weakens the underlying antiferromagnetic correlations in NCCO. 
Still, our measurements indicate that a majority of the dynamic correlations at \qco~are magnetic in nature.

Overall, our findings further support the scenario in which the static CO in NCCO originates from the pinning of dynamic correlations with decreasing temperature. 
Indeed, in YBCO, the pinning of static CO correlations by native defects is supported by nuclear magnetic resonance measurements \cite{Wu_2015}, while the CO correlation length in \hg~appears to be limited by a specific disorder potential \cite{Tabis_PRB}. Detailed spatially-resolved measurements will be necessary to identify the role of disorder in the pinning of CO correlations in NCCO.
Upon cooling, we observed a softening of the electronic response, which, according to a recent theoretical prediction \cite{Greco_PRB}, may be the first signature of a $d$-wave CO in NCCO. This observation may have implications for the mechanism of superconductivity in the electron-doped cuprates: a quantum Monte Carlo study showed that a $d$-wave CO at the measured \qco~implies the presence of nematic fluctuations that also enhance $d$-wave superconductivity \cite{Li_2017}.
Finally, our systematic investigation revealed a coupling between dynamic magnetic and charge-order correlations, which might also be present in other hole- and electron-doped cuprates. Depending on the strength of the effect, a similar coupling might be detectable in those materials by future RIXS studies following the methodology described here.

\begin{acknowledgments}
We thank the ESRF for the allocation of synchrotron radiation beamtime. We acknowledge G.\,Ghiringhelli, L.\,Braicovich, R. Fumagalli and E.\,Weschke for fruitful discussions. This research was undertaken thanks in part to funding from the Max Planck-UBC-UTokyo Centre for Quantum Materials and the Canada First Research Excellence Fund, Quantum Materials and Future Technologies Program. The work at UBC was supported by the Killam, Alfred P. Sloan, and Natural Sciences and Engineering Research Council of Canada’s (NSERC\rq{s}) Steacie Memorial Fellowships (A.D.); the Alexander von Humboldt Fellowship (A.D.); the Canada Research Chairs Program (A.D.); and the NSERC, Canada Foundation for Innovation (CFI), and CIFAR Quantum Materials. The work at the University of Minnesota was supported by the NSF through the University of Minnesota Materials Research Science and Engineering Center under award no.\,DMR-1420013.
\end{acknowledgments}

\bibliographystyle{apsrev4-1}
\bibliography{bib_lib}

\begin{thebibliography}{52}%
\makeatletter
\providecommand \@ifxundefined [1]{%
 \@ifx{#1\undefined}
}%
\providecommand \@ifnum [1]{%
 \ifnum #1\expandafter \@firstoftwo
 \else \expandafter \@secondoftwo
 \fi
}%
\providecommand \@ifx [1]{%
 \ifx #1\expandafter \@firstoftwo
 \else \expandafter \@secondoftwo
 \fi
}%
\providecommand \natexlab [1]{#1}%
\providecommand \enquote  [1]{``#1''}%
\providecommand \bibnamefont  [1]{#1}%
\providecommand \bibfnamefont [1]{#1}%
\providecommand \citenamefont [1]{#1}%
\providecommand \href@noop [0]{\@secondoftwo}%
\providecommand \href [0]{\begingroup \@sanitize@url \@href}%
\providecommand \@href[1]{\@@startlink{#1}\@@href}%
\providecommand \@@href[1]{\endgroup#1\@@endlink}%
\providecommand \@sanitize@url [0]{\catcode `\\12\catcode `\$12\catcode
  `\&12\catcode `\#12\catcode `\^12\catcode `\_12\catcode `\%12\relax}%
\providecommand \@@startlink[1]{}%
\providecommand \@@endlink[0]{}%
\providecommand \url  [0]{\begingroup\@sanitize@url \@url }%
\providecommand \@url [1]{\endgroup\@href {#1}{\urlprefix }}%
\providecommand \urlprefix  [0]{URL }%
\providecommand \Eprint [0]{\href }%
\providecommand \doibase [0]{http://dx.doi.org/}%
\providecommand \selectlanguage [0]{\@gobble}%
\providecommand \bibinfo  [0]{\@secondoftwo}%
\providecommand \bibfield  [0]{\@secondoftwo}%
\providecommand \translation [1]{[#1]}%
\providecommand \BibitemOpen [0]{}%
\providecommand \bibitemStop [0]{}%
\providecommand \bibitemNoStop [0]{.\EOS\space}%
\providecommand \EOS [0]{\spacefactor3000\relax}%
\providecommand \BibitemShut  [1]{\csname bibitem#1\endcsname}%
\let\auto@bib@innerbib\@empty
\bibitem [{\citenamefont {Tranquada}\ \emph {et~al.}(1995)\citenamefont
  {Tranquada}, \citenamefont {Sternlieb}, \citenamefont {Axe}, \citenamefont
  {Nakamura},\ and\ \citenamefont {Uchida}}]{tranquada_evidence_1995}%
  \BibitemOpen
  \bibfield  {author} {\bibinfo {author} {\bibfnamefont {J.~M.}\ \bibnamefont
  {Tranquada}}, \bibinfo {author} {\bibfnamefont {B.~J.}\ \bibnamefont
  {Sternlieb}}, \bibinfo {author} {\bibfnamefont {J.~D.}\ \bibnamefont {Axe}},
  \bibinfo {author} {\bibfnamefont {Y.}~\bibnamefont {Nakamura}}, \ and\
  \bibinfo {author} {\bibfnamefont {S.}~\bibnamefont {Uchida}},\ }\href
  {\doibase 10.1038/375561a0} {\bibfield  {journal} {\bibinfo  {journal}
  {Nature}\ }\textbf {\bibinfo {volume} {375}},\ \bibinfo {pages} {561}
  (\bibinfo {year} {1995})}\BibitemShut {NoStop}%
\bibitem [{\citenamefont {Hoffman}\ \emph {et~al.}(2002)\citenamefont
  {Hoffman}, \citenamefont {Hudson}, \citenamefont {Lang}, \citenamefont
  {Madhavan}, \citenamefont {Eisaki}, \citenamefont {Uchida},\ and\
  \citenamefont {Davis}}]{Hoffman_2002}%
  \BibitemOpen
  \bibfield  {author} {\bibinfo {author} {\bibfnamefont {J.~E.}\ \bibnamefont
  {Hoffman}}, \bibinfo {author} {\bibfnamefont {E.~W.}\ \bibnamefont {Hudson}},
  \bibinfo {author} {\bibfnamefont {K.~M.}\ \bibnamefont {Lang}}, \bibinfo
  {author} {\bibfnamefont {V.}~\bibnamefont {Madhavan}}, \bibinfo {author}
  {\bibfnamefont {H.}~\bibnamefont {Eisaki}}, \bibinfo {author} {\bibfnamefont
  {S.}~\bibnamefont {Uchida}}, \ and\ \bibinfo {author} {\bibfnamefont {J.~C.}\
  \bibnamefont {Davis}},\ }\href {\doibase 10.1126/science.1066974} {\bibfield
  {journal} {\bibinfo  {journal} {Science}\ }\textbf {\bibinfo {volume}
  {295}},\ \bibinfo {pages} {466} (\bibinfo {year} {2002})}\BibitemShut
  {NoStop}%
\bibitem [{\citenamefont {Howald}\ \emph {et~al.}(2003)\citenamefont {Howald},
  \citenamefont {Eisaki}, \citenamefont {Kaneko},\ and\ \citenamefont
  {Kapitulnik}}]{Howald_2003}%
  \BibitemOpen
  \bibfield  {author} {\bibinfo {author} {\bibfnamefont {C.}~\bibnamefont
  {Howald}}, \bibinfo {author} {\bibfnamefont {H.}~\bibnamefont {Eisaki}},
  \bibinfo {author} {\bibfnamefont {N.}~\bibnamefont {Kaneko}}, \ and\ \bibinfo
  {author} {\bibfnamefont {A.}~\bibnamefont {Kapitulnik}},\ }\href {\doibase
  10.1073/pnas.1233768100} {\bibfield  {journal} {\bibinfo  {journal}
  {Proceedings of the National Academy of Sciences}\ }\textbf {\bibinfo
  {volume} {100}},\ \bibinfo {pages} {9705} (\bibinfo {year}
  {2003})}\BibitemShut {NoStop}%
\bibitem [{\citenamefont {{Abbamonte}}\ \emph {et~al.}(2005)\citenamefont
  {{Abbamonte}}, \citenamefont {{Rusydi}}, \citenamefont {{Smadici}},
  \citenamefont {{Gu}}, \citenamefont {{Sawatzky}},\ and\ \citenamefont
  {{Feng}}}]{Abbamonte_2005}%
  \BibitemOpen
  \bibfield  {author} {\bibinfo {author} {\bibfnamefont {P.}~\bibnamefont
  {{Abbamonte}}}, \bibinfo {author} {\bibfnamefont {A.}~\bibnamefont
  {{Rusydi}}}, \bibinfo {author} {\bibfnamefont {S.}~\bibnamefont {{Smadici}}},
  \bibinfo {author} {\bibfnamefont {G.~D.}\ \bibnamefont {{Gu}}}, \bibinfo
  {author} {\bibfnamefont {G.~A.}\ \bibnamefont {{Sawatzky}}}, \ and\ \bibinfo
  {author} {\bibfnamefont {D.~L.}\ \bibnamefont {{Feng}}},\ }\href {\doibase
  10.1038/nphys178} {\bibfield  {journal} {\bibinfo  {journal} {Nature
  Physics}\ }\textbf {\bibinfo {volume} {1}},\ \bibinfo {pages} {155} (\bibinfo
  {year} {2005})}\BibitemShut {NoStop}%
\bibitem [{\citenamefont {Wu}\ \emph {et~al.}(2011)\citenamefont {Wu},
  \citenamefont {Mayaffre}, \citenamefont {Kramer}, \citenamefont {Horvatic},
  \citenamefont {Berthier}, \citenamefont {Hardy}, \citenamefont {Liang},
  \citenamefont {Bonn},\ and\ \citenamefont {Julien}}]{Wu_2011}%
  \BibitemOpen
  \bibfield  {author} {\bibinfo {author} {\bibfnamefont {T.}~\bibnamefont
  {Wu}}, \bibinfo {author} {\bibfnamefont {H.}~\bibnamefont {Mayaffre}},
  \bibinfo {author} {\bibfnamefont {S.}~\bibnamefont {Kramer}}, \bibinfo
  {author} {\bibfnamefont {M.}~\bibnamefont {Horvatic}}, \bibinfo {author}
  {\bibfnamefont {C.}~\bibnamefont {Berthier}}, \bibinfo {author}
  {\bibfnamefont {W.~N.}\ \bibnamefont {Hardy}}, \bibinfo {author}
  {\bibfnamefont {R.}~\bibnamefont {Liang}}, \bibinfo {author} {\bibfnamefont
  {D.~A.}\ \bibnamefont {Bonn}}, \ and\ \bibinfo {author} {\bibfnamefont
  {M.-H.}\ \bibnamefont {Julien}},\ }\href@noop {} {\bibfield  {journal}
  {\bibinfo  {journal} {Nature}\ }\textbf {\bibinfo {volume} {477}},\ \bibinfo
  {pages} {191} (\bibinfo {year} {2011})}\BibitemShut {NoStop}%
\bibitem [{\citenamefont {Ghiringhelli}\ \emph {et~al.}(2012)\citenamefont
  {Ghiringhelli}, \citenamefont {Le~Tacon}, \citenamefont {Minola},
  \citenamefont {Blanco-Canosa}, \citenamefont {Mazzoli}, \citenamefont
  {Brookes}, \citenamefont {De~Luca}, \citenamefont {Frano}, \citenamefont
  {Hawthorn}, \citenamefont {He}, \citenamefont {Loew}, \citenamefont {Sala},
  \citenamefont {Peets}, \citenamefont {Salluzzo}, \citenamefont {Schierle},
  \citenamefont {Sutarto}, \citenamefont {Sawatzky}, \citenamefont {Weschke},
  \citenamefont {Keimer},\ and\ \citenamefont
  {Braicovich}}]{Ghiringhelli_CDW_2012}%
  \BibitemOpen
  \bibfield  {author} {\bibinfo {author} {\bibfnamefont {G.}~\bibnamefont
  {Ghiringhelli}}, \bibinfo {author} {\bibfnamefont {M.}~\bibnamefont
  {Le~Tacon}}, \bibinfo {author} {\bibfnamefont {M.}~\bibnamefont {Minola}},
  \bibinfo {author} {\bibfnamefont {S.}~\bibnamefont {Blanco-Canosa}}, \bibinfo
  {author} {\bibfnamefont {C.}~\bibnamefont {Mazzoli}}, \bibinfo {author}
  {\bibfnamefont {N.~B.}\ \bibnamefont {Brookes}}, \bibinfo {author}
  {\bibfnamefont {G.~M.}\ \bibnamefont {De~Luca}}, \bibinfo {author}
  {\bibfnamefont {A.}~\bibnamefont {Frano}}, \bibinfo {author} {\bibfnamefont
  {D.~G.}\ \bibnamefont {Hawthorn}}, \bibinfo {author} {\bibfnamefont
  {F.}~\bibnamefont {He}}, \bibinfo {author} {\bibfnamefont {T.}~\bibnamefont
  {Loew}}, \bibinfo {author} {\bibfnamefont {M.~M.}\ \bibnamefont {Sala}},
  \bibinfo {author} {\bibfnamefont {D.~C.}\ \bibnamefont {Peets}}, \bibinfo
  {author} {\bibfnamefont {M.}~\bibnamefont {Salluzzo}}, \bibinfo {author}
  {\bibfnamefont {E.}~\bibnamefont {Schierle}}, \bibinfo {author}
  {\bibfnamefont {R.}~\bibnamefont {Sutarto}}, \bibinfo {author} {\bibfnamefont
  {G.~A.}\ \bibnamefont {Sawatzky}}, \bibinfo {author} {\bibfnamefont
  {E.}~\bibnamefont {Weschke}}, \bibinfo {author} {\bibfnamefont
  {B.}~\bibnamefont {Keimer}}, \ and\ \bibinfo {author} {\bibfnamefont
  {L.}~\bibnamefont {Braicovich}},\ }\href {\doibase 10.1126/science.1223532}
  {\bibfield  {journal} {\bibinfo  {journal} {Science}\ }\textbf {\bibinfo
  {volume} {337}},\ \bibinfo {pages} {821} (\bibinfo {year}
  {2012})}\BibitemShut {NoStop}%
\bibitem [{\citenamefont {Chang}\ \emph {et~al.}(2012)\citenamefont {Chang},
  \citenamefont {Blackburn}, \citenamefont {Holmes}, \citenamefont
  {Christensen}, \citenamefont {Larsen}, \citenamefont {Mesot}, \citenamefont
  {Liang}, \citenamefont {Bonn}, \citenamefont {Hardy}, \citenamefont
  {Watenphul}, \citenamefont {Zimmermann}, \citenamefont {Forgan},\ and\
  \citenamefont {Hayden}}]{Chang_CDW_2012}%
  \BibitemOpen
  \bibfield  {author} {\bibinfo {author} {\bibfnamefont {J.}~\bibnamefont
  {Chang}}, \bibinfo {author} {\bibfnamefont {E.}~\bibnamefont {Blackburn}},
  \bibinfo {author} {\bibfnamefont {A.~T.}\ \bibnamefont {Holmes}}, \bibinfo
  {author} {\bibfnamefont {N.~B.}\ \bibnamefont {Christensen}}, \bibinfo
  {author} {\bibfnamefont {J.}~\bibnamefont {Larsen}}, \bibinfo {author}
  {\bibfnamefont {J.}~\bibnamefont {Mesot}}, \bibinfo {author} {\bibfnamefont
  {R.}~\bibnamefont {Liang}}, \bibinfo {author} {\bibfnamefont {D.~A.}\
  \bibnamefont {Bonn}}, \bibinfo {author} {\bibfnamefont {W.~N.}\ \bibnamefont
  {Hardy}}, \bibinfo {author} {\bibfnamefont {A.}~\bibnamefont {Watenphul}},
  \bibinfo {author} {\bibfnamefont {M.~v.}\ \bibnamefont {Zimmermann}},
  \bibinfo {author} {\bibfnamefont {E.~M.}\ \bibnamefont {Forgan}}, \ and\
  \bibinfo {author} {\bibfnamefont {S.~M.}\ \bibnamefont {Hayden}},\
  }\href@noop {} {\bibfield  {journal} {\bibinfo  {journal} {Nature Physics}\
  }\textbf {\bibinfo {volume} {8}},\ \bibinfo {pages} {871} (\bibinfo {year}
  {2012})}\BibitemShut {NoStop}%
\bibitem [{\citenamefont {Achkar}\ \emph {et~al.}(2012)\citenamefont {Achkar},
  \citenamefont {Sutarto}, \citenamefont {Mao}, \citenamefont {He},
  \citenamefont {Frano}, \citenamefont {Blanco-Canosa}, \citenamefont
  {Le~Tacon}, \citenamefont {Ghiringhelli}, \citenamefont {Braicovich},
  \citenamefont {Minola}, \citenamefont {Moretti~Sala}, \citenamefont
  {Mazzoli}, \citenamefont {Liang}, \citenamefont {Bonn}, \citenamefont
  {Hardy}, \citenamefont {Keimer}, \citenamefont {Sawatzky},\ and\
  \citenamefont {Hawthorn}}]{Achkar_2012}%
  \BibitemOpen
  \bibfield  {author} {\bibinfo {author} {\bibfnamefont {A.~J.}\ \bibnamefont
  {Achkar}}, \bibinfo {author} {\bibfnamefont {R.}~\bibnamefont {Sutarto}},
  \bibinfo {author} {\bibfnamefont {X.}~\bibnamefont {Mao}}, \bibinfo {author}
  {\bibfnamefont {F.}~\bibnamefont {He}}, \bibinfo {author} {\bibfnamefont
  {A.}~\bibnamefont {Frano}}, \bibinfo {author} {\bibfnamefont
  {S.}~\bibnamefont {Blanco-Canosa}}, \bibinfo {author} {\bibfnamefont
  {M.}~\bibnamefont {Le~Tacon}}, \bibinfo {author} {\bibfnamefont
  {G.}~\bibnamefont {Ghiringhelli}}, \bibinfo {author} {\bibfnamefont
  {L.}~\bibnamefont {Braicovich}}, \bibinfo {author} {\bibfnamefont
  {M.}~\bibnamefont {Minola}}, \bibinfo {author} {\bibfnamefont
  {M.}~\bibnamefont {Moretti~Sala}}, \bibinfo {author} {\bibfnamefont
  {C.}~\bibnamefont {Mazzoli}}, \bibinfo {author} {\bibfnamefont
  {R.}~\bibnamefont {Liang}}, \bibinfo {author} {\bibfnamefont {D.~A.}\
  \bibnamefont {Bonn}}, \bibinfo {author} {\bibfnamefont {W.~N.}\ \bibnamefont
  {Hardy}}, \bibinfo {author} {\bibfnamefont {B.}~\bibnamefont {Keimer}},
  \bibinfo {author} {\bibfnamefont {G.~A.}\ \bibnamefont {Sawatzky}}, \ and\
  \bibinfo {author} {\bibfnamefont {D.~G.}\ \bibnamefont {Hawthorn}},\ }\href
  {\doibase 10.1103/PhysRevLett.109.167001} {\bibfield  {journal} {\bibinfo
  {journal} {Phys. Rev. Lett.}\ }\textbf {\bibinfo {volume} {109}},\ \bibinfo
  {pages} {167001} (\bibinfo {year} {2012})}\BibitemShut {NoStop}%
\bibitem [{\citenamefont {Comin}\ \emph {et~al.}(2014)\citenamefont {Comin},
  \citenamefont {Frano}, \citenamefont {Yee}, \citenamefont {Yoshida},
  \citenamefont {Eisaki}, \citenamefont {Schierle}, \citenamefont {Weschke},
  \citenamefont {Sutarto}, \citenamefont {He}, \citenamefont {Soumyanarayanan},
  \citenamefont {He}, \citenamefont {Le~Tacon}, \citenamefont {Elfimov},
  \citenamefont {Hoffman}, \citenamefont {Sawatzky}, \citenamefont {Keimer},\
  and\ \citenamefont {Damascelli}}]{Comin_CDW_2014}%
  \BibitemOpen
  \bibfield  {author} {\bibinfo {author} {\bibfnamefont {R.}~\bibnamefont
  {Comin}}, \bibinfo {author} {\bibfnamefont {A.}~\bibnamefont {Frano}},
  \bibinfo {author} {\bibfnamefont {M.~M.}\ \bibnamefont {Yee}}, \bibinfo
  {author} {\bibfnamefont {Y.}~\bibnamefont {Yoshida}}, \bibinfo {author}
  {\bibfnamefont {H.}~\bibnamefont {Eisaki}}, \bibinfo {author} {\bibfnamefont
  {E.}~\bibnamefont {Schierle}}, \bibinfo {author} {\bibfnamefont
  {E.}~\bibnamefont {Weschke}}, \bibinfo {author} {\bibfnamefont
  {R.}~\bibnamefont {Sutarto}}, \bibinfo {author} {\bibfnamefont
  {F.}~\bibnamefont {He}}, \bibinfo {author} {\bibfnamefont {A.}~\bibnamefont
  {Soumyanarayanan}}, \bibinfo {author} {\bibfnamefont {Y.}~\bibnamefont {He}},
  \bibinfo {author} {\bibfnamefont {M.}~\bibnamefont {Le~Tacon}}, \bibinfo
  {author} {\bibfnamefont {I.~S.}\ \bibnamefont {Elfimov}}, \bibinfo {author}
  {\bibfnamefont {J.~E.}\ \bibnamefont {Hoffman}}, \bibinfo {author}
  {\bibfnamefont {G.~A.}\ \bibnamefont {Sawatzky}}, \bibinfo {author}
  {\bibfnamefont {B.}~\bibnamefont {Keimer}}, \ and\ \bibinfo {author}
  {\bibfnamefont {A.}~\bibnamefont {Damascelli}},\ }\href {\doibase
  10.1126/science.1242996} {\bibfield  {journal} {\bibinfo  {journal}
  {Science}\ }\textbf {\bibinfo {volume} {343}},\ \bibinfo {pages} {390}
  (\bibinfo {year} {2014})}\BibitemShut {NoStop}%
\bibitem [{\citenamefont {da~Silva~Neto}\ \emph {et~al.}(2014)\citenamefont
  {da~Silva~Neto}, \citenamefont {Aynajian}, \citenamefont {Frano},
  \citenamefont {Comin}, \citenamefont {Schierle}, \citenamefont {Weschke},
  \citenamefont {Gyenis}, \citenamefont {Wen}, \citenamefont {Schneeloch},
  \citenamefont {Xu}, \citenamefont {Ono}, \citenamefont {Gu}, \citenamefont
  {Le~Tacon},\ and\ \citenamefont {Yazdani}}]{daSilvaNeto_CDW_2014}%
  \BibitemOpen
  \bibfield  {author} {\bibinfo {author} {\bibfnamefont {E.~H.}\ \bibnamefont
  {da~Silva~Neto}}, \bibinfo {author} {\bibfnamefont {P.}~\bibnamefont
  {Aynajian}}, \bibinfo {author} {\bibfnamefont {A.}~\bibnamefont {Frano}},
  \bibinfo {author} {\bibfnamefont {R.}~\bibnamefont {Comin}}, \bibinfo
  {author} {\bibfnamefont {E.}~\bibnamefont {Schierle}}, \bibinfo {author}
  {\bibfnamefont {E.}~\bibnamefont {Weschke}}, \bibinfo {author} {\bibfnamefont
  {A.}~\bibnamefont {Gyenis}}, \bibinfo {author} {\bibfnamefont
  {J.}~\bibnamefont {Wen}}, \bibinfo {author} {\bibfnamefont {J.}~\bibnamefont
  {Schneeloch}}, \bibinfo {author} {\bibfnamefont {Z.}~\bibnamefont {Xu}},
  \bibinfo {author} {\bibfnamefont {S.}~\bibnamefont {Ono}}, \bibinfo {author}
  {\bibfnamefont {G.}~\bibnamefont {Gu}}, \bibinfo {author} {\bibfnamefont
  {M.}~\bibnamefont {Le~Tacon}}, \ and\ \bibinfo {author} {\bibfnamefont
  {A.}~\bibnamefont {Yazdani}},\ }\href {\doibase 10.1126/science.1243479}
  {\bibfield  {journal} {\bibinfo  {journal} {Science}\ }\textbf {\bibinfo
  {volume} {343}},\ \bibinfo {pages} {393} (\bibinfo {year}
  {2014})}\BibitemShut {NoStop}%
\bibitem [{\citenamefont {Hashimoto}\ \emph {et~al.}(2014)\citenamefont
  {Hashimoto}, \citenamefont {Ghiringhelli}, \citenamefont {Lee}, \citenamefont
  {Dellea}, \citenamefont {Amorese}, \citenamefont {Mazzoli}, \citenamefont
  {Kummer}, \citenamefont {Brookes}, \citenamefont {Moritz}, \citenamefont
  {Yoshida}, \citenamefont {Eisaki}, \citenamefont {Hussain}, \citenamefont
  {Devereaux}, \citenamefont {Shen},\ and\ \citenamefont
  {Braicovich}}]{Hashimoto_2014}%
  \BibitemOpen
  \bibfield  {author} {\bibinfo {author} {\bibfnamefont {M.}~\bibnamefont
  {Hashimoto}}, \bibinfo {author} {\bibfnamefont {G.}~\bibnamefont
  {Ghiringhelli}}, \bibinfo {author} {\bibfnamefont {W.-S.}\ \bibnamefont
  {Lee}}, \bibinfo {author} {\bibfnamefont {G.}~\bibnamefont {Dellea}},
  \bibinfo {author} {\bibfnamefont {A.}~\bibnamefont {Amorese}}, \bibinfo
  {author} {\bibfnamefont {C.}~\bibnamefont {Mazzoli}}, \bibinfo {author}
  {\bibfnamefont {K.}~\bibnamefont {Kummer}}, \bibinfo {author} {\bibfnamefont
  {N.~B.}\ \bibnamefont {Brookes}}, \bibinfo {author} {\bibfnamefont
  {B.}~\bibnamefont {Moritz}}, \bibinfo {author} {\bibfnamefont
  {Y.}~\bibnamefont {Yoshida}}, \bibinfo {author} {\bibfnamefont
  {H.}~\bibnamefont {Eisaki}}, \bibinfo {author} {\bibfnamefont
  {Z.}~\bibnamefont {Hussain}}, \bibinfo {author} {\bibfnamefont {T.~P.}\
  \bibnamefont {Devereaux}}, \bibinfo {author} {\bibfnamefont {Z.-X.}\
  \bibnamefont {Shen}}, \ and\ \bibinfo {author} {\bibfnamefont
  {L.}~\bibnamefont {Braicovich}},\ }\href {\doibase
  10.1103/PhysRevB.89.220511} {\bibfield  {journal} {\bibinfo  {journal} {Phys.
  Rev. B}\ }\textbf {\bibinfo {volume} {89}},\ \bibinfo {pages} {220511}
  (\bibinfo {year} {2014})}\BibitemShut {NoStop}%
\bibitem [{\citenamefont {{Tabis}}\ \emph {et~al.}(2014)\citenamefont
  {{Tabis}}, \citenamefont {{Li}}, \citenamefont {{Le Tacon}}, \citenamefont
  {{Braicovich}}, \citenamefont {{Kreyssig}}, \citenamefont {{Minola}},
  \citenamefont {{Dellea}}, \citenamefont {{Weschke}}, \citenamefont {{Veit}},
  \citenamefont {{Ramazanoglu}}, \citenamefont {{Goldman}}, \citenamefont
  {{Schmitt}}, \citenamefont {{Ghiringhelli}}, \citenamefont {{Bari{\v
  s}i{\'c}}}, \citenamefont {{Chan}}, \citenamefont {{Dorow}}, \citenamefont
  {{Yu}}, \citenamefont {{Zhao}}, \citenamefont {{Keimer}},\ and\ \citenamefont
  {{Greven}}}]{Tabis_2014}%
  \BibitemOpen
  \bibfield  {author} {\bibinfo {author} {\bibfnamefont {W.}~\bibnamefont
  {{Tabis}}}, \bibinfo {author} {\bibfnamefont {Y.}~\bibnamefont {{Li}}},
  \bibinfo {author} {\bibfnamefont {M.}~\bibnamefont {{Le Tacon}}}, \bibinfo
  {author} {\bibfnamefont {L.}~\bibnamefont {{Braicovich}}}, \bibinfo {author}
  {\bibfnamefont {A.}~\bibnamefont {{Kreyssig}}}, \bibinfo {author}
  {\bibfnamefont {M.}~\bibnamefont {{Minola}}}, \bibinfo {author}
  {\bibfnamefont {G.}~\bibnamefont {{Dellea}}}, \bibinfo {author}
  {\bibfnamefont {E.}~\bibnamefont {{Weschke}}}, \bibinfo {author}
  {\bibfnamefont {M.~J.}\ \bibnamefont {{Veit}}}, \bibinfo {author}
  {\bibfnamefont {M.}~\bibnamefont {{Ramazanoglu}}}, \bibinfo {author}
  {\bibfnamefont {A.~I.}\ \bibnamefont {{Goldman}}}, \bibinfo {author}
  {\bibfnamefont {T.}~\bibnamefont {{Schmitt}}}, \bibinfo {author}
  {\bibfnamefont {G.}~\bibnamefont {{Ghiringhelli}}}, \bibinfo {author}
  {\bibfnamefont {N.}~\bibnamefont {{Bari{\v s}i{\'c}}}}, \bibinfo {author}
  {\bibfnamefont {M.~K.}\ \bibnamefont {{Chan}}}, \bibinfo {author}
  {\bibfnamefont {C.~J.}\ \bibnamefont {{Dorow}}}, \bibinfo {author}
  {\bibfnamefont {G.}~\bibnamefont {{Yu}}}, \bibinfo {author} {\bibfnamefont
  {X.}~\bibnamefont {{Zhao}}}, \bibinfo {author} {\bibfnamefont
  {B.}~\bibnamefont {{Keimer}}}, \ and\ \bibinfo {author} {\bibfnamefont
  {M.}~\bibnamefont {{Greven}}},\ }\href@noop {} {\bibfield  {journal}
  {\bibinfo  {journal} {Nat. Comm.}\ }\textbf {\bibinfo {volume} {5}} (\bibinfo
  {year} {2014})}\BibitemShut {NoStop}%
\bibitem [{\citenamefont {da~Silva~Neto}\ \emph {et~al.}(2015)\citenamefont
  {da~Silva~Neto}, \citenamefont {Comin}, \citenamefont {He}, \citenamefont
  {Sutarto}, \citenamefont {Jiang}, \citenamefont {Greene}, \citenamefont
  {Sawatzky},\ and\ \citenamefont {Damascelli}}]{daSilvaNeto_Science_2015}%
  \BibitemOpen
  \bibfield  {author} {\bibinfo {author} {\bibfnamefont {E.~H.}\ \bibnamefont
  {da~Silva~Neto}}, \bibinfo {author} {\bibfnamefont {R.}~\bibnamefont
  {Comin}}, \bibinfo {author} {\bibfnamefont {F.}~\bibnamefont {He}}, \bibinfo
  {author} {\bibfnamefont {R.}~\bibnamefont {Sutarto}}, \bibinfo {author}
  {\bibfnamefont {Y.}~\bibnamefont {Jiang}}, \bibinfo {author} {\bibfnamefont
  {R.~L.}\ \bibnamefont {Greene}}, \bibinfo {author} {\bibfnamefont {G.~A.}\
  \bibnamefont {Sawatzky}}, \ and\ \bibinfo {author} {\bibfnamefont
  {A.}~\bibnamefont {Damascelli}},\ }\href
  {http://www.sciencemag.org/content/347/6219/282.abstract N2 - In cuprate
  high-temperature superconductors, an antiferromagnetic Mott insulating state
  can be destabilized toward unconventional superconductivity by either hole or
  electron doping. In hole-doped (p-type) cuprates, a charge ordering (CO)
  instability competes with superconductivity inside the pseudogap state. We
  report resonant x-ray scattering measurements that demonstrate the presence
  of charge ordering in the n-type cuprate Nd2–xCexCuO4 near optimal doping.
  We find that the CO in Nd2–xCexCuO4 occurs with similar periodicity, and
  along the same direction, as in p-type cuprates. However, in contrast to the
  latter, the CO onset in Nd2–xCexCuO4 is higher than the pseudogap
  temperature, and is in the temperature range where antiferromagnetic
  fluctuations are first detected. Our discovery opens a parallel path to the
  study of CO and its relationship to antiferromagnetism and
  superconductivity.} {\bibfield  {journal} {\bibinfo  {journal} {Science}\
  }\textbf {\bibinfo {volume} {347}},\ \bibinfo {pages} {282} (\bibinfo {year}
  {2015})}\BibitemShut {NoStop}%
\bibitem [{\citenamefont {da~Silva~Neto}\ \emph {et~al.}(2016)\citenamefont
  {da~Silva~Neto}, \citenamefont {Yu}, \citenamefont {Minola}, \citenamefont
  {Sutarto}, \citenamefont {Schierle}, \citenamefont {Boschini}, \citenamefont
  {Zonno}, \citenamefont {Bluschke}, \citenamefont {Higgins}, \citenamefont
  {Li}, \citenamefont {Yu}, \citenamefont {Weschke}, \citenamefont {He},
  \citenamefont {Le~Tacon}, \citenamefont {Greene}, \citenamefont {Greven},
  \citenamefont {Sawatzky}, \citenamefont {Keimer},\ and\ \citenamefont
  {Damascelli}}]{daSilvaNeto_SciAdv_2016}%
  \BibitemOpen
  \bibfield  {author} {\bibinfo {author} {\bibfnamefont {E.~H.}\ \bibnamefont
  {da~Silva~Neto}}, \bibinfo {author} {\bibfnamefont {B.}~\bibnamefont {Yu}},
  \bibinfo {author} {\bibfnamefont {M.}~\bibnamefont {Minola}}, \bibinfo
  {author} {\bibfnamefont {R.}~\bibnamefont {Sutarto}}, \bibinfo {author}
  {\bibfnamefont {E.}~\bibnamefont {Schierle}}, \bibinfo {author}
  {\bibfnamefont {F.}~\bibnamefont {Boschini}}, \bibinfo {author}
  {\bibfnamefont {M.}~\bibnamefont {Zonno}}, \bibinfo {author} {\bibfnamefont
  {M.}~\bibnamefont {Bluschke}}, \bibinfo {author} {\bibfnamefont
  {J.}~\bibnamefont {Higgins}}, \bibinfo {author} {\bibfnamefont
  {Y.}~\bibnamefont {Li}}, \bibinfo {author} {\bibfnamefont {G.}~\bibnamefont
  {Yu}}, \bibinfo {author} {\bibfnamefont {E.}~\bibnamefont {Weschke}},
  \bibinfo {author} {\bibfnamefont {F.}~\bibnamefont {He}}, \bibinfo {author}
  {\bibfnamefont {M.}~\bibnamefont {Le~Tacon}}, \bibinfo {author}
  {\bibfnamefont {R.~L.}\ \bibnamefont {Greene}}, \bibinfo {author}
  {\bibfnamefont {M.}~\bibnamefont {Greven}}, \bibinfo {author} {\bibfnamefont
  {G.~A.}\ \bibnamefont {Sawatzky}}, \bibinfo {author} {\bibfnamefont
  {B.}~\bibnamefont {Keimer}}, \ and\ \bibinfo {author} {\bibfnamefont
  {A.}~\bibnamefont {Damascelli}},\ }\href
  {http://advances.sciencemag.org/content/2/8/e1600782.abstract} {\bibfield
  {journal} {\bibinfo  {journal} {Science Advances}\ }\textbf {\bibinfo
  {volume} {2}} (\bibinfo {year} {2016})}\BibitemShut {NoStop}%
\bibitem [{\citenamefont {Tabis}\ \emph {et~al.}(2017)\citenamefont {Tabis},
  \citenamefont {Yu}, \citenamefont {Bialo}, \citenamefont {Bluschke},
  \citenamefont {Kolodziej}, \citenamefont {Kozlowski}, \citenamefont
  {Blackburn}, \citenamefont {Sen}, \citenamefont {Forgan}, \citenamefont
  {Zimmermann}, \citenamefont {Tang}, \citenamefont {Weschke}, \citenamefont
  {Vignolle}, \citenamefont {Hepting}, \citenamefont {Gretarsson},
  \citenamefont {Sutarto}, \citenamefont {He}, \citenamefont {Le~Tacon},
  \citenamefont {Bari\ifmmode \check{s}\else \v{s}\fi{}i\ifmmode~\acute{c}\else
  \'{c}\fi{}}, \citenamefont {Yu},\ and\ \citenamefont {Greven}}]{Tabis_PRB}%
  \BibitemOpen
  \bibfield  {author} {\bibinfo {author} {\bibfnamefont {W.}~\bibnamefont
  {Tabis}}, \bibinfo {author} {\bibfnamefont {B.}~\bibnamefont {Yu}}, \bibinfo
  {author} {\bibfnamefont {I.}~\bibnamefont {Bialo}}, \bibinfo {author}
  {\bibfnamefont {M.}~\bibnamefont {Bluschke}}, \bibinfo {author}
  {\bibfnamefont {T.}~\bibnamefont {Kolodziej}}, \bibinfo {author}
  {\bibfnamefont {A.}~\bibnamefont {Kozlowski}}, \bibinfo {author}
  {\bibfnamefont {E.}~\bibnamefont {Blackburn}}, \bibinfo {author}
  {\bibfnamefont {K.}~\bibnamefont {Sen}}, \bibinfo {author} {\bibfnamefont
  {E.~M.}\ \bibnamefont {Forgan}}, \bibinfo {author} {\bibfnamefont {M.~v.}\
  \bibnamefont {Zimmermann}}, \bibinfo {author} {\bibfnamefont
  {Y.}~\bibnamefont {Tang}}, \bibinfo {author} {\bibfnamefont {E.}~\bibnamefont
  {Weschke}}, \bibinfo {author} {\bibfnamefont {B.}~\bibnamefont {Vignolle}},
  \bibinfo {author} {\bibfnamefont {M.}~\bibnamefont {Hepting}}, \bibinfo
  {author} {\bibfnamefont {H.}~\bibnamefont {Gretarsson}}, \bibinfo {author}
  {\bibfnamefont {R.}~\bibnamefont {Sutarto}}, \bibinfo {author} {\bibfnamefont
  {F.}~\bibnamefont {He}}, \bibinfo {author} {\bibfnamefont {M.}~\bibnamefont
  {Le~Tacon}}, \bibinfo {author} {\bibfnamefont {N.}~\bibnamefont {Bari\ifmmode
  \check{s}\else \v{s}\fi{}i\ifmmode~\acute{c}\else \'{c}\fi{}}}, \bibinfo
  {author} {\bibfnamefont {G.}~\bibnamefont {Yu}}, \ and\ \bibinfo {author}
  {\bibfnamefont {M.}~\bibnamefont {Greven}},\ }\href {\doibase
  10.1103/PhysRevB.96.134510} {\bibfield  {journal} {\bibinfo  {journal} {Phys.
  Rev. B}\ }\textbf {\bibinfo {volume} {96}},\ \bibinfo {pages} {134510}
  (\bibinfo {year} {2017})}\BibitemShut {NoStop}%
\bibitem [{\citenamefont {Jang}\ \emph {et~al.}(2017)\citenamefont {Jang},
  \citenamefont {Asano}, \citenamefont {Fujita}, \citenamefont {Hashimoto},
  \citenamefont {Lu}, \citenamefont {Burns}, \citenamefont {Kao},\ and\
  \citenamefont {Lee}}]{PRX_NCCO}%
  \BibitemOpen
  \bibfield  {author} {\bibinfo {author} {\bibfnamefont {H.}~\bibnamefont
  {Jang}}, \bibinfo {author} {\bibfnamefont {S.}~\bibnamefont {Asano}},
  \bibinfo {author} {\bibfnamefont {M.}~\bibnamefont {Fujita}}, \bibinfo
  {author} {\bibfnamefont {M.}~\bibnamefont {Hashimoto}}, \bibinfo {author}
  {\bibfnamefont {D.~H.}\ \bibnamefont {Lu}}, \bibinfo {author} {\bibfnamefont
  {C.~A.}\ \bibnamefont {Burns}}, \bibinfo {author} {\bibfnamefont {C.-C.}\
  \bibnamefont {Kao}}, \ and\ \bibinfo {author} {\bibfnamefont {J.-S.}\
  \bibnamefont {Lee}},\ }\href {\doibase 10.1103/PhysRevX.7.041066} {\bibfield
  {journal} {\bibinfo  {journal} {Phys. Rev. X}\ }\textbf {\bibinfo {volume}
  {7}},\ \bibinfo {pages} {041066} (\bibinfo {year} {2017})}\BibitemShut
  {NoStop}%
\bibitem [{\citenamefont {Zaanen}\ and\ \citenamefont
  {Gunnarsson}(1989)}]{zaanen_charged_1989}%
  \BibitemOpen
  \bibfield  {author} {\bibinfo {author} {\bibfnamefont {J.}~\bibnamefont
  {Zaanen}}\ and\ \bibinfo {author} {\bibfnamefont {O.}~\bibnamefont
  {Gunnarsson}},\ }\href {\doibase 10.1103/PhysRevB.40.7391} {\bibfield
  {journal} {\bibinfo  {journal} {Phys. Rev. B}\ }\textbf {\bibinfo {volume}
  {40}},\ \bibinfo {pages} {7391} (\bibinfo {year} {1989})}\BibitemShut
  {NoStop}%
\bibitem [{\citenamefont {Kivelson}\ \emph {et~al.}(1998)\citenamefont
  {Kivelson}, \citenamefont {Fradkin},\ and\ \citenamefont
  {Emery}}]{Kivelson1998}%
  \BibitemOpen
  \bibfield  {author} {\bibinfo {author} {\bibfnamefont {S.~A.}\ \bibnamefont
  {Kivelson}}, \bibinfo {author} {\bibfnamefont {E.}~\bibnamefont {Fradkin}}, \
  and\ \bibinfo {author} {\bibfnamefont {V.~J.}\ \bibnamefont {Emery}},\ }\href
  {http://dx.doi.org/10.1038/31177} {\bibfield  {journal} {\bibinfo  {journal}
  {Nature}\ }\textbf {\bibinfo {volume} {393}},\ \bibinfo {pages} {550}
  (\bibinfo {year} {1998})}\BibitemShut {NoStop}%
\bibitem [{\citenamefont {Davis}\ and\ \citenamefont
  {Lee}(2013)}]{Davis_DHLee_2013}%
  \BibitemOpen
  \bibfield  {author} {\bibinfo {author} {\bibfnamefont {J.~C.~S.}\
  \bibnamefont {Davis}}\ and\ \bibinfo {author} {\bibfnamefont {D.-H.}\
  \bibnamefont {Lee}},\ }\href {\doibase 10.1073/pnas.1316512110} {\bibfield
  {journal} {\bibinfo  {journal} {Proceedings of the National Academy of
  Sciences}\ }\textbf {\bibinfo {volume} {110}},\ \bibinfo {pages} {17623}
  (\bibinfo {year} {2013})}\BibitemShut {NoStop}%
\bibitem [{\citenamefont {Efetov}\ \emph {et~al.}(2013)\citenamefont {Efetov},
  \citenamefont {Meier},\ and\ \citenamefont {Pepin}}]{Efetov_nat_phys_2013}%
  \BibitemOpen
  \bibfield  {author} {\bibinfo {author} {\bibfnamefont {K.~B.}\ \bibnamefont
  {Efetov}}, \bibinfo {author} {\bibfnamefont {H.}~\bibnamefont {Meier}}, \
  and\ \bibinfo {author} {\bibfnamefont {C.}~\bibnamefont {Pepin}},\ }\href
  {http://dx.doi.org/10.1038/nphys2641} {\bibfield  {journal} {\bibinfo
  {journal} {Nat Phys}\ }\textbf {\bibinfo {volume} {9}},\ \bibinfo {pages}
  {442} (\bibinfo {year} {2013})}\BibitemShut {NoStop}%
\bibitem [{\citenamefont {Sachdev}\ and\ \citenamefont
  {La~Placa}(2013)}]{Sachdev_PRL_2013}%
  \BibitemOpen
  \bibfield  {author} {\bibinfo {author} {\bibfnamefont {S.}~\bibnamefont
  {Sachdev}}\ and\ \bibinfo {author} {\bibfnamefont {R.}~\bibnamefont
  {La~Placa}},\ }\href {\doibase 10.1103/PhysRevLett.111.027202} {\bibfield
  {journal} {\bibinfo  {journal} {Phys. Rev. Lett.}\ }\textbf {\bibinfo
  {volume} {111}},\ \bibinfo {pages} {027202} (\bibinfo {year}
  {2013})}\BibitemShut {NoStop}%
\bibitem [{\citenamefont {Wang}\ and\ \citenamefont
  {Chubukov}(2014)}]{Wang_Chubukov_PRB_2014}%
  \BibitemOpen
  \bibfield  {author} {\bibinfo {author} {\bibfnamefont {Y.}~\bibnamefont
  {Wang}}\ and\ \bibinfo {author} {\bibfnamefont {A.}~\bibnamefont
  {Chubukov}},\ }\href {\doibase 10.1103/PhysRevB.90.035149} {\bibfield
  {journal} {\bibinfo  {journal} {Phys. Rev. B}\ }\textbf {\bibinfo {volume}
  {90}},\ \bibinfo {pages} {035149} (\bibinfo {year} {2014})}\BibitemShut
  {NoStop}%
\bibitem [{\citenamefont {Blanco-Canosa}\ \emph {et~al.}(2014)\citenamefont
  {Blanco-Canosa}, \citenamefont {Frano}, \citenamefont {Schierle},
  \citenamefont {Porras}, \citenamefont {Loew}, \citenamefont {Minola},
  \citenamefont {Bluschke}, \citenamefont {Weschke}, \citenamefont {Keimer},\
  and\ \citenamefont {Le~Tacon}}]{Santi_2014}%
  \BibitemOpen
  \bibfield  {author} {\bibinfo {author} {\bibfnamefont {S.}~\bibnamefont
  {Blanco-Canosa}}, \bibinfo {author} {\bibfnamefont {A.}~\bibnamefont
  {Frano}}, \bibinfo {author} {\bibfnamefont {E.}~\bibnamefont {Schierle}},
  \bibinfo {author} {\bibfnamefont {J.}~\bibnamefont {Porras}}, \bibinfo
  {author} {\bibfnamefont {T.}~\bibnamefont {Loew}}, \bibinfo {author}
  {\bibfnamefont {M.}~\bibnamefont {Minola}}, \bibinfo {author} {\bibfnamefont
  {M.}~\bibnamefont {Bluschke}}, \bibinfo {author} {\bibfnamefont
  {E.}~\bibnamefont {Weschke}}, \bibinfo {author} {\bibfnamefont
  {B.}~\bibnamefont {Keimer}}, \ and\ \bibinfo {author} {\bibfnamefont
  {M.}~\bibnamefont {Le~Tacon}},\ }\href {\doibase 10.1103/PhysRevB.90.054513}
  {\bibfield  {journal} {\bibinfo  {journal} {Phys. Rev. B}\ }\textbf {\bibinfo
  {volume} {90}},\ \bibinfo {pages} {054513} (\bibinfo {year}
  {2014})}\BibitemShut {NoStop}%
\bibitem [{\citenamefont {H\"ucker}\ \emph {et~al.}(2014)\citenamefont
  {H\"ucker}, \citenamefont {Christensen}, \citenamefont {Holmes},
  \citenamefont {Blackburn}, \citenamefont {Forgan}, \citenamefont {Liang},
  \citenamefont {Bonn}, \citenamefont {Hardy}, \citenamefont {Gutowski},
  \citenamefont {Zimmermann}, \citenamefont {Hayden},\ and\ \citenamefont
  {Chang}}]{Hucker_PhysRevB.90.054514}%
  \BibitemOpen
  \bibfield  {author} {\bibinfo {author} {\bibfnamefont {M.}~\bibnamefont
  {H\"ucker}}, \bibinfo {author} {\bibfnamefont {N.~B.}\ \bibnamefont
  {Christensen}}, \bibinfo {author} {\bibfnamefont {A.~T.}\ \bibnamefont
  {Holmes}}, \bibinfo {author} {\bibfnamefont {E.}~\bibnamefont {Blackburn}},
  \bibinfo {author} {\bibfnamefont {E.~M.}\ \bibnamefont {Forgan}}, \bibinfo
  {author} {\bibfnamefont {R.}~\bibnamefont {Liang}}, \bibinfo {author}
  {\bibfnamefont {D.~A.}\ \bibnamefont {Bonn}}, \bibinfo {author}
  {\bibfnamefont {W.~N.}\ \bibnamefont {Hardy}}, \bibinfo {author}
  {\bibfnamefont {O.}~\bibnamefont {Gutowski}}, \bibinfo {author}
  {\bibfnamefont {M.~v.}\ \bibnamefont {Zimmermann}}, \bibinfo {author}
  {\bibfnamefont {S.~M.}\ \bibnamefont {Hayden}}, \ and\ \bibinfo {author}
  {\bibfnamefont {J.}~\bibnamefont {Chang}},\ }\href {\doibase
  10.1103/PhysRevB.90.054514} {\bibfield  {journal} {\bibinfo  {journal} {Phys.
  Rev. B}\ }\textbf {\bibinfo {volume} {90}},\ \bibinfo {pages} {054514}
  (\bibinfo {year} {2014})}\BibitemShut {NoStop}%
\bibitem [{\citenamefont {Blanco-Canosa}\ \emph {et~al.}(2013)\citenamefont
  {Blanco-Canosa}, \citenamefont {Frano}, \citenamefont {Loew}, \citenamefont
  {Lu}, \citenamefont {Porras}, \citenamefont {Ghiringhelli}, \citenamefont
  {Minola}, \citenamefont {Mazzoli}, \citenamefont {Braicovich}, \citenamefont
  {Schierle}, \citenamefont {Weschke}, \citenamefont {{Le Tacon}},\ and\
  \citenamefont {Keimer}}]{Santi_PRL2013}%
  \BibitemOpen
  \bibfield  {author} {\bibinfo {author} {\bibfnamefont {S.}~\bibnamefont
  {Blanco-Canosa}}, \bibinfo {author} {\bibfnamefont {A.}~\bibnamefont
  {Frano}}, \bibinfo {author} {\bibfnamefont {T.}~\bibnamefont {Loew}},
  \bibinfo {author} {\bibfnamefont {Y.}~\bibnamefont {Lu}}, \bibinfo {author}
  {\bibfnamefont {J.}~\bibnamefont {Porras}}, \bibinfo {author} {\bibfnamefont
  {G.}~\bibnamefont {Ghiringhelli}}, \bibinfo {author} {\bibfnamefont
  {M.}~\bibnamefont {Minola}}, \bibinfo {author} {\bibfnamefont
  {C.}~\bibnamefont {Mazzoli}}, \bibinfo {author} {\bibfnamefont
  {L.}~\bibnamefont {Braicovich}}, \bibinfo {author} {\bibfnamefont
  {E.}~\bibnamefont {Schierle}}, \bibinfo {author} {\bibfnamefont
  {E.}~\bibnamefont {Weschke}}, \bibinfo {author} {\bibfnamefont
  {M.}~\bibnamefont {{Le Tacon}}}, \ and\ \bibinfo {author} {\bibfnamefont
  {B.}~\bibnamefont {Keimer}},\ }\href {\doibase
  10.1103/PhysRevLett.110.187001} {\bibfield  {journal} {\bibinfo  {journal}
  {Phys. Rev. Lett.}\ }\textbf {\bibinfo {volume} {110}},\ \bibinfo {pages} {1}
  (\bibinfo {year} {2013})}\BibitemShut {NoStop}%
\bibitem [{\citenamefont {{Motoyama}}\ \emph {et~al.}(2007)\citenamefont
  {{Motoyama}}, \citenamefont {{Yu}}, \citenamefont {{Vishik}}, \citenamefont
  {{Vajk}}, \citenamefont {{Mang}},\ and\ \citenamefont
  {{Greven}}}]{Motoyama_Spin_2007}%
  \BibitemOpen
  \bibfield  {author} {\bibinfo {author} {\bibfnamefont {E.~M.}\ \bibnamefont
  {{Motoyama}}}, \bibinfo {author} {\bibfnamefont {G.}~\bibnamefont {{Yu}}},
  \bibinfo {author} {\bibfnamefont {I.~M.}\ \bibnamefont {{Vishik}}}, \bibinfo
  {author} {\bibfnamefont {O.~P.}\ \bibnamefont {{Vajk}}}, \bibinfo {author}
  {\bibfnamefont {P.~K.}\ \bibnamefont {{Mang}}}, \ and\ \bibinfo {author}
  {\bibfnamefont {M.}~\bibnamefont {{Greven}}},\ }\href {\doibase
  10.1038/nature05437} {\bibfield  {journal} {\bibinfo  {journal} {Nature}\
  }\textbf {\bibinfo {volume} {445}},\ \bibinfo {pages} {186} (\bibinfo {year}
  {2007})}\BibitemShut {NoStop}%
\bibitem [{\citenamefont {Le~Tacon}\ \emph {et~al.}(2014)\citenamefont
  {Le~Tacon}, \citenamefont {Bosak}, \citenamefont {Souliou}, \citenamefont
  {Dellea}, \citenamefont {Loew}, \citenamefont {Heid}, \citenamefont {Bohnen},
  \citenamefont {Ghiringhelli}, \citenamefont {Krisch},\ and\ \citenamefont
  {Keimer}}]{LeTacon2014}%
  \BibitemOpen
  \bibfield  {author} {\bibinfo {author} {\bibfnamefont {M.}~\bibnamefont
  {Le~Tacon}}, \bibinfo {author} {\bibfnamefont {A.}~\bibnamefont {Bosak}},
  \bibinfo {author} {\bibfnamefont {S.~M.}\ \bibnamefont {Souliou}}, \bibinfo
  {author} {\bibfnamefont {G.}~\bibnamefont {Dellea}}, \bibinfo {author}
  {\bibfnamefont {T.}~\bibnamefont {Loew}}, \bibinfo {author} {\bibfnamefont
  {R.}~\bibnamefont {Heid}}, \bibinfo {author} {\bibfnamefont {K.-P.}\
  \bibnamefont {Bohnen}}, \bibinfo {author} {\bibfnamefont {G.}~\bibnamefont
  {Ghiringhelli}}, \bibinfo {author} {\bibfnamefont {M.}~\bibnamefont
  {Krisch}}, \ and\ \bibinfo {author} {\bibfnamefont {B.}~\bibnamefont
  {Keimer}},\ }\href {http://dx.doi.org/10.1038/nphys2805} {\bibfield
  {journal} {\bibinfo  {journal} {Nat Phys}\ }\textbf {\bibinfo {volume}
  {10}},\ \bibinfo {pages} {52} (\bibinfo {year} {2014})}\BibitemShut {NoStop}%
\bibitem [{\citenamefont {Wu}\ \emph {et~al.}(2015)\citenamefont {Wu},
  \citenamefont {Mayaffre}, \citenamefont {Kr{\"a}mer}, \citenamefont
  {Horvati{\'c}}, \citenamefont {Berthier}, \citenamefont {Hardy},
  \citenamefont {Liang}, \citenamefont {Bonn},\ and\ \citenamefont
  {Julien}}]{Wu_2015}%
  \BibitemOpen
  \bibfield  {author} {\bibinfo {author} {\bibfnamefont {T.}~\bibnamefont
  {Wu}}, \bibinfo {author} {\bibfnamefont {H.}~\bibnamefont {Mayaffre}},
  \bibinfo {author} {\bibfnamefont {S.}~\bibnamefont {Kr{\"a}mer}}, \bibinfo
  {author} {\bibfnamefont {M.}~\bibnamefont {Horvati{\'c}}}, \bibinfo {author}
  {\bibfnamefont {C.}~\bibnamefont {Berthier}}, \bibinfo {author}
  {\bibfnamefont {W.~N.}\ \bibnamefont {Hardy}}, \bibinfo {author}
  {\bibfnamefont {R.}~\bibnamefont {Liang}}, \bibinfo {author} {\bibfnamefont
  {D.~A.}\ \bibnamefont {Bonn}}, \ and\ \bibinfo {author} {\bibfnamefont
  {M.-H.}\ \bibnamefont {Julien}},\ }\href@noop {} {\bibfield  {journal}
  {\bibinfo  {journal} {Nat Commun}\ }\textbf {\bibinfo {volume} {6}} (\bibinfo
  {year} {2015})}\BibitemShut {NoStop}%
\bibitem [{\citenamefont {Gerber}\ \emph {et~al.}(2015)\citenamefont {Gerber},
  \citenamefont {Jang}, \citenamefont {Nojiri}, \citenamefont {Matsuzawa},
  \citenamefont {Yasumura}, \citenamefont {Bonn}, \citenamefont {Liang},
  \citenamefont {Hardy}, \citenamefont {Islam}, \citenamefont {Mehta},
  \citenamefont {Song}, \citenamefont {Sikorski}, \citenamefont {Stefanescu},
  \citenamefont {Feng}, \citenamefont {Kivelson}, \citenamefont {Devereaux},
  \citenamefont {Shen}, \citenamefont {Kao}, \citenamefont {Lee}, \citenamefont
  {Zhu},\ and\ \citenamefont {Lee}}]{Gerber_2015}%
  \BibitemOpen
  \bibfield  {author} {\bibinfo {author} {\bibfnamefont {S.}~\bibnamefont
  {Gerber}}, \bibinfo {author} {\bibfnamefont {H.}~\bibnamefont {Jang}},
  \bibinfo {author} {\bibfnamefont {H.}~\bibnamefont {Nojiri}}, \bibinfo
  {author} {\bibfnamefont {S.}~\bibnamefont {Matsuzawa}}, \bibinfo {author}
  {\bibfnamefont {H.}~\bibnamefont {Yasumura}}, \bibinfo {author}
  {\bibfnamefont {D.~A.}\ \bibnamefont {Bonn}}, \bibinfo {author}
  {\bibfnamefont {R.}~\bibnamefont {Liang}}, \bibinfo {author} {\bibfnamefont
  {W.~N.}\ \bibnamefont {Hardy}}, \bibinfo {author} {\bibfnamefont
  {Z.}~\bibnamefont {Islam}}, \bibinfo {author} {\bibfnamefont
  {A.}~\bibnamefont {Mehta}}, \bibinfo {author} {\bibfnamefont
  {S.}~\bibnamefont {Song}}, \bibinfo {author} {\bibfnamefont {M.}~\bibnamefont
  {Sikorski}}, \bibinfo {author} {\bibfnamefont {D.}~\bibnamefont
  {Stefanescu}}, \bibinfo {author} {\bibfnamefont {Y.}~\bibnamefont {Feng}},
  \bibinfo {author} {\bibfnamefont {S.~A.}\ \bibnamefont {Kivelson}}, \bibinfo
  {author} {\bibfnamefont {T.~P.}\ \bibnamefont {Devereaux}}, \bibinfo {author}
  {\bibfnamefont {Z.-X.}\ \bibnamefont {Shen}}, \bibinfo {author}
  {\bibfnamefont {C.-C.}\ \bibnamefont {Kao}}, \bibinfo {author} {\bibfnamefont
  {W.-S.}\ \bibnamefont {Lee}}, \bibinfo {author} {\bibfnamefont
  {D.}~\bibnamefont {Zhu}}, \ and\ \bibinfo {author} {\bibfnamefont {J.-S.}\
  \bibnamefont {Lee}},\ }\href@noop {} {\bibfield  {journal} {\bibinfo
  {journal} {Science}\ }\textbf {\bibinfo {volume} {350}},\ \bibinfo {pages}
  {949} (\bibinfo {year} {2015})}\BibitemShut {NoStop}%
\bibitem [{\citenamefont {Chang}\ \emph {et~al.}(2016)\citenamefont {Chang},
  \citenamefont {Blackburn}, \citenamefont {Ivashko}, \citenamefont {Holmes},
  \citenamefont {Christensen}, \citenamefont {Hücker}, \citenamefont {Liang},
  \citenamefont {Bonn}, \citenamefont {Hardy}, \citenamefont {Rütt},
  \citenamefont {Zimmermann}, \citenamefont {Forgan},\ and\ \citenamefont
  {Hayden}}]{Chang2016}%
  \BibitemOpen
  \bibfield  {author} {\bibinfo {author} {\bibfnamefont {J.}~\bibnamefont
  {Chang}}, \bibinfo {author} {\bibfnamefont {E.}~\bibnamefont {Blackburn}},
  \bibinfo {author} {\bibfnamefont {O.}~\bibnamefont {Ivashko}}, \bibinfo
  {author} {\bibfnamefont {A.~T.}\ \bibnamefont {Holmes}}, \bibinfo {author}
  {\bibfnamefont {N.~B.}\ \bibnamefont {Christensen}}, \bibinfo {author}
  {\bibfnamefont {M.}~\bibnamefont {Hücker}}, \bibinfo {author} {\bibfnamefont
  {R.}~\bibnamefont {Liang}}, \bibinfo {author} {\bibfnamefont {D.~A.}\
  \bibnamefont {Bonn}}, \bibinfo {author} {\bibfnamefont {W.~N.}\ \bibnamefont
  {Hardy}}, \bibinfo {author} {\bibfnamefont {U.}~\bibnamefont {Rütt}},
  \bibinfo {author} {\bibfnamefont {M.~v.}\ \bibnamefont {Zimmermann}},
  \bibinfo {author} {\bibfnamefont {E.~M.}\ \bibnamefont {Forgan}}, \ and\
  \bibinfo {author} {\bibfnamefont {S.~M.}\ \bibnamefont {Hayden}},\ }\href
  {http://dx.doi.org/10.1038/ncomms11494} {\bibfield  {journal} {\bibinfo
  {journal} {Nature Communications}\ }\textbf {\bibinfo {volume} {7}},\
  \bibinfo {pages} {11494} (\bibinfo {year} {2016})}\BibitemShut {NoStop}%
\bibitem [{\citenamefont {Jang}\ \emph {et~al.}(2016)\citenamefont {Jang},
  \citenamefont {Lee}, \citenamefont {Nojiri}, \citenamefont {Matsuzawa},
  \citenamefont {Yasumura}, \citenamefont {Nie}, \citenamefont {Maharaj},
  \citenamefont {Gerber}, \citenamefont {Liu}, \citenamefont {Mehta},
  \citenamefont {Bonn}, \citenamefont {Liang}, \citenamefont {Hardy},
  \citenamefont {Burns}, \citenamefont {Islam}, \citenamefont {Song},
  \citenamefont {Hastings}, \citenamefont {Devereaux}, \citenamefont {Shen},
  \citenamefont {Kivelson}, \citenamefont {Kao}, \citenamefont {Zhu},\ and\
  \citenamefont {Lee}}]{Jang2016}%
  \BibitemOpen
  \bibfield  {author} {\bibinfo {author} {\bibfnamefont {H.}~\bibnamefont
  {Jang}}, \bibinfo {author} {\bibfnamefont {W.-S.}\ \bibnamefont {Lee}},
  \bibinfo {author} {\bibfnamefont {H.}~\bibnamefont {Nojiri}}, \bibinfo
  {author} {\bibfnamefont {S.}~\bibnamefont {Matsuzawa}}, \bibinfo {author}
  {\bibfnamefont {H.}~\bibnamefont {Yasumura}}, \bibinfo {author}
  {\bibfnamefont {L.}~\bibnamefont {Nie}}, \bibinfo {author} {\bibfnamefont
  {A.~V.}\ \bibnamefont {Maharaj}}, \bibinfo {author} {\bibfnamefont
  {S.}~\bibnamefont {Gerber}}, \bibinfo {author} {\bibfnamefont {Y.-J.}\
  \bibnamefont {Liu}}, \bibinfo {author} {\bibfnamefont {A.}~\bibnamefont
  {Mehta}}, \bibinfo {author} {\bibfnamefont {D.~A.}\ \bibnamefont {Bonn}},
  \bibinfo {author} {\bibfnamefont {R.}~\bibnamefont {Liang}}, \bibinfo
  {author} {\bibfnamefont {W.~N.}\ \bibnamefont {Hardy}}, \bibinfo {author}
  {\bibfnamefont {C.~A.}\ \bibnamefont {Burns}}, \bibinfo {author}
  {\bibfnamefont {Z.}~\bibnamefont {Islam}}, \bibinfo {author} {\bibfnamefont
  {S.}~\bibnamefont {Song}}, \bibinfo {author} {\bibfnamefont {J.}~\bibnamefont
  {Hastings}}, \bibinfo {author} {\bibfnamefont {T.~P.}\ \bibnamefont
  {Devereaux}}, \bibinfo {author} {\bibfnamefont {Z.-X.}\ \bibnamefont {Shen}},
  \bibinfo {author} {\bibfnamefont {S.~A.}\ \bibnamefont {Kivelson}}, \bibinfo
  {author} {\bibfnamefont {C.-C.}\ \bibnamefont {Kao}}, \bibinfo {author}
  {\bibfnamefont {D.}~\bibnamefont {Zhu}}, \ and\ \bibinfo {author}
  {\bibfnamefont {J.-S.}\ \bibnamefont {Lee}},\ }\href
  {http://www.pnas.org/content/early/2016/11/29/1612849113.abstract} {\bibfield
   {journal} {\bibinfo  {journal} {Proceedings of the National Academy of
  Sciences}\ }\textbf {\bibinfo {volume} {113}},\ \bibinfo {pages} {14645}
  (\bibinfo {year} {2016})}\BibitemShut {NoStop}%
\bibitem [{\citenamefont {Miao}\ \emph {et~al.}(2017)\citenamefont {Miao},
  \citenamefont {Lorenzana}, \citenamefont {Seibold}, \citenamefont {Peng},
  \citenamefont {Amorese}, \citenamefont {Yakhou-Harris}, \citenamefont
  {Kummer}, \citenamefont {Brookes}, \citenamefont {Konik}, \citenamefont
  {Thampy}, \citenamefont {Gu}, \citenamefont {Ghiringhelli}, \citenamefont
  {Braicovich},\ and\ \citenamefont {Dean}}]{Miao_2017}%
  \BibitemOpen
  \bibfield  {author} {\bibinfo {author} {\bibfnamefont {H.}~\bibnamefont
  {Miao}}, \bibinfo {author} {\bibfnamefont {J.}~\bibnamefont {Lorenzana}},
  \bibinfo {author} {\bibfnamefont {G.}~\bibnamefont {Seibold}}, \bibinfo
  {author} {\bibfnamefont {Y.~Y.}\ \bibnamefont {Peng}}, \bibinfo {author}
  {\bibfnamefont {A.}~\bibnamefont {Amorese}}, \bibinfo {author} {\bibfnamefont
  {F.}~\bibnamefont {Yakhou-Harris}}, \bibinfo {author} {\bibfnamefont
  {K.}~\bibnamefont {Kummer}}, \bibinfo {author} {\bibfnamefont {N.~B.}\
  \bibnamefont {Brookes}}, \bibinfo {author} {\bibfnamefont {R.~M.}\
  \bibnamefont {Konik}}, \bibinfo {author} {\bibfnamefont {V.}~\bibnamefont
  {Thampy}}, \bibinfo {author} {\bibfnamefont {G.~D.}\ \bibnamefont {Gu}},
  \bibinfo {author} {\bibfnamefont {G.}~\bibnamefont {Ghiringhelli}}, \bibinfo
  {author} {\bibfnamefont {L.}~\bibnamefont {Braicovich}}, \ and\ \bibinfo
  {author} {\bibfnamefont {M.~P.~M.}\ \bibnamefont {Dean}},\ }\href {\doibase
  10.1073/pnas.1708549114} {\bibfield  {journal} {\bibinfo  {journal}
  {Proceedings of the National Academy of Sciences}\ }\textbf {\bibinfo
  {volume} {114}},\ \bibinfo {pages} {12430} (\bibinfo {year}
  {2017})}\BibitemShut {NoStop}%
\bibitem [{\citenamefont {Kivelson}\ \emph {et~al.}(2003)\citenamefont
  {Kivelson}, \citenamefont {Bindloss}, \citenamefont {Fradkin}, \citenamefont
  {Oganesyan}, \citenamefont {Tranquada}, \citenamefont {Kapitulnik},\ and\
  \citenamefont {Howald}}]{Kivelson_RMP_2002}%
  \BibitemOpen
  \bibfield  {author} {\bibinfo {author} {\bibfnamefont {S.~A.}\ \bibnamefont
  {Kivelson}}, \bibinfo {author} {\bibfnamefont {I.~P.}\ \bibnamefont
  {Bindloss}}, \bibinfo {author} {\bibfnamefont {E.}~\bibnamefont {Fradkin}},
  \bibinfo {author} {\bibfnamefont {V.}~\bibnamefont {Oganesyan}}, \bibinfo
  {author} {\bibfnamefont {J.~M.}\ \bibnamefont {Tranquada}}, \bibinfo {author}
  {\bibfnamefont {A.}~\bibnamefont {Kapitulnik}}, \ and\ \bibinfo {author}
  {\bibfnamefont {C.}~\bibnamefont {Howald}},\ }\href {\doibase
  10.1103/RevModPhys.75.1201} {\bibfield  {journal} {\bibinfo  {journal} {Rev.
  Mod. Phys.}\ }\textbf {\bibinfo {volume} {75}},\ \bibinfo {pages} {1201}
  (\bibinfo {year} {2003})}\BibitemShut {NoStop}%
\bibitem [{\citenamefont {Caplan}\ \emph {et~al.}(2015)\citenamefont {Caplan},
  \citenamefont {Wachtel},\ and\ \citenamefont {Orgad}}]{Caplan_PRB_2015}%
  \BibitemOpen
  \bibfield  {author} {\bibinfo {author} {\bibfnamefont {Y.}~\bibnamefont
  {Caplan}}, \bibinfo {author} {\bibfnamefont {G.}~\bibnamefont {Wachtel}}, \
  and\ \bibinfo {author} {\bibfnamefont {D.}~\bibnamefont {Orgad}},\ }\href
  {\doibase 10.1103/PhysRevB.92.224504} {\bibfield  {journal} {\bibinfo
  {journal} {Phys. Rev. B}\ }\textbf {\bibinfo {volume} {92}},\ \bibinfo
  {pages} {224504} (\bibinfo {year} {2015})}\BibitemShut {NoStop}%
\bibitem [{\citenamefont {Wu}\ \emph {et~al.}(2016)\citenamefont {Wu},
  \citenamefont {Zhou}, \citenamefont {Hirata}, \citenamefont {Vinograd},
  \citenamefont {Mayaffre}, \citenamefont {Liang}, \citenamefont {Hardy},
  \citenamefont {Bonn}, \citenamefont {Loew}, \citenamefont {Porras},
  \citenamefont {Haug}, \citenamefont {Lin}, \citenamefont {Hinkov},
  \citenamefont {Keimer},\ and\ \citenamefont {Julien}}]{Wu_PRB_2016}%
  \BibitemOpen
  \bibfield  {author} {\bibinfo {author} {\bibfnamefont {T.}~\bibnamefont
  {Wu}}, \bibinfo {author} {\bibfnamefont {R.}~\bibnamefont {Zhou}}, \bibinfo
  {author} {\bibfnamefont {M.}~\bibnamefont {Hirata}}, \bibinfo {author}
  {\bibfnamefont {I.}~\bibnamefont {Vinograd}}, \bibinfo {author}
  {\bibfnamefont {H.}~\bibnamefont {Mayaffre}}, \bibinfo {author}
  {\bibfnamefont {R.}~\bibnamefont {Liang}}, \bibinfo {author} {\bibfnamefont
  {W.~N.}\ \bibnamefont {Hardy}}, \bibinfo {author} {\bibfnamefont {D.~A.}\
  \bibnamefont {Bonn}}, \bibinfo {author} {\bibfnamefont {T.}~\bibnamefont
  {Loew}}, \bibinfo {author} {\bibfnamefont {J.}~\bibnamefont {Porras}},
  \bibinfo {author} {\bibfnamefont {D.}~\bibnamefont {Haug}}, \bibinfo {author}
  {\bibfnamefont {C.~T.}\ \bibnamefont {Lin}}, \bibinfo {author} {\bibfnamefont
  {V.}~\bibnamefont {Hinkov}}, \bibinfo {author} {\bibfnamefont
  {B.}~\bibnamefont {Keimer}}, \ and\ \bibinfo {author} {\bibfnamefont {M.-H.}\
  \bibnamefont {Julien}},\ }\href {\doibase 10.1103/PhysRevB.93.134518}
  {\bibfield  {journal} {\bibinfo  {journal} {Phys. Rev. B}\ }\textbf {\bibinfo
  {volume} {93}},\ \bibinfo {pages} {134518} (\bibinfo {year}
  {2016})}\BibitemShut {NoStop}%
\bibitem [{\citenamefont {Chaix}\ \emph {et~al.}(2017)\citenamefont {Chaix},
  \citenamefont {Ghiringhelli}, \citenamefont {Peng}, \citenamefont
  {Hashimoto}, \citenamefont {Moritz}, \citenamefont {Kummer}, \citenamefont
  {Brookes}, \citenamefont {He}, \citenamefont {Chen}, \citenamefont {Ishida},
  \citenamefont {Yoshida}, \citenamefont {Eisaki}, \citenamefont {Salluzzo},
  \citenamefont {Braicovich}, \citenamefont {Shen}, \citenamefont {Devereaux},\
  and\ \citenamefont {Lee}}]{Chaix2017}%
  \BibitemOpen
  \bibfield  {author} {\bibinfo {author} {\bibfnamefont {L.}~\bibnamefont
  {Chaix}}, \bibinfo {author} {\bibfnamefont {G.}~\bibnamefont {Ghiringhelli}},
  \bibinfo {author} {\bibfnamefont {Y.~Y.}\ \bibnamefont {Peng}}, \bibinfo
  {author} {\bibfnamefont {M.}~\bibnamefont {Hashimoto}}, \bibinfo {author}
  {\bibfnamefont {B.}~\bibnamefont {Moritz}}, \bibinfo {author} {\bibfnamefont
  {K.}~\bibnamefont {Kummer}}, \bibinfo {author} {\bibfnamefont {N.~B.}\
  \bibnamefont {Brookes}}, \bibinfo {author} {\bibfnamefont {Y.}~\bibnamefont
  {He}}, \bibinfo {author} {\bibfnamefont {S.}~\bibnamefont {Chen}}, \bibinfo
  {author} {\bibfnamefont {S.}~\bibnamefont {Ishida}}, \bibinfo {author}
  {\bibfnamefont {Y.}~\bibnamefont {Yoshida}}, \bibinfo {author} {\bibfnamefont
  {H.}~\bibnamefont {Eisaki}}, \bibinfo {author} {\bibfnamefont
  {M.}~\bibnamefont {Salluzzo}}, \bibinfo {author} {\bibfnamefont
  {L.}~\bibnamefont {Braicovich}}, \bibinfo {author} {\bibfnamefont {Z.-X.}\
  \bibnamefont {Shen}}, \bibinfo {author} {\bibfnamefont {T.~P.}\ \bibnamefont
  {Devereaux}}, \ and\ \bibinfo {author} {\bibfnamefont {W.-S.}\ \bibnamefont
  {Lee}},\ }\href {http://dx.doi.org/10.1038/nphys4157} {\bibfield  {journal}
  {\bibinfo  {journal} {Nature Physics}\ }\textbf {\bibinfo {volume} {13}},\
  \bibinfo {pages} {952} (\bibinfo {year} {2017})}\BibitemShut {NoStop}%
\bibitem [{\citenamefont {Thampy}\ \emph {et~al.}(2014)\citenamefont {Thampy},
  \citenamefont {Dean}, \citenamefont {Christensen}, \citenamefont {Steinke},
  \citenamefont {Islam}, \citenamefont {Oda}, \citenamefont {Ido},
  \citenamefont {Momono}, \citenamefont {Wilkins},\ and\ \citenamefont
  {Hill}}]{Thampy_2014}%
  \BibitemOpen
  \bibfield  {author} {\bibinfo {author} {\bibfnamefont {V.}~\bibnamefont
  {Thampy}}, \bibinfo {author} {\bibfnamefont {M.~P.~M.}\ \bibnamefont {Dean}},
  \bibinfo {author} {\bibfnamefont {N.~B.}\ \bibnamefont {Christensen}},
  \bibinfo {author} {\bibfnamefont {L.}~\bibnamefont {Steinke}}, \bibinfo
  {author} {\bibfnamefont {Z.}~\bibnamefont {Islam}}, \bibinfo {author}
  {\bibfnamefont {M.}~\bibnamefont {Oda}}, \bibinfo {author} {\bibfnamefont
  {M.}~\bibnamefont {Ido}}, \bibinfo {author} {\bibfnamefont {N.}~\bibnamefont
  {Momono}}, \bibinfo {author} {\bibfnamefont {S.~B.}\ \bibnamefont {Wilkins}},
  \ and\ \bibinfo {author} {\bibfnamefont {J.~P.}\ \bibnamefont {Hill}},\
  }\href {\doibase 10.1103/PhysRevB.90.100510} {\bibfield  {journal} {\bibinfo
  {journal} {Phys. Rev. B}\ }\textbf {\bibinfo {volume} {90}},\ \bibinfo
  {pages} {100510} (\bibinfo {year} {2014})}\BibitemShut {NoStop}%
\bibitem [{SM()}]{SM}%
  \BibitemOpen
  \href@noop {} {\ }\bibinfo {note} {See supplemental materials.}\BibitemShut
  {Stop}%
\bibitem [{\citenamefont {Sala}\ \emph {et~al.}(2011)\citenamefont {Sala},
  \citenamefont {Bisogni}, \citenamefont {Aruta}, \citenamefont {Balestrino},
  \citenamefont {Berger}, \citenamefont {Brookes}, \citenamefont {de~Luca},
  \citenamefont {Castro}, \citenamefont {Grioni}, \citenamefont {Guarise},
  \citenamefont {Medaglia}, \citenamefont {Granozio}, \citenamefont {Minola},
  \citenamefont {Perna}, \citenamefont {Radovic}, \citenamefont {Salluzzo},
  \citenamefont {Schmitt}, \citenamefont {Zhou}, \citenamefont {Braicovich},\
  and\ \citenamefont {Ghiringhelli}}]{MMS_dd}%
  \BibitemOpen
  \bibfield  {author} {\bibinfo {author} {\bibfnamefont {M.~M.}\ \bibnamefont
  {Sala}}, \bibinfo {author} {\bibfnamefont {V.}~\bibnamefont {Bisogni}},
  \bibinfo {author} {\bibfnamefont {C.}~\bibnamefont {Aruta}}, \bibinfo
  {author} {\bibfnamefont {G.}~\bibnamefont {Balestrino}}, \bibinfo {author}
  {\bibfnamefont {H.}~\bibnamefont {Berger}}, \bibinfo {author} {\bibfnamefont
  {N.~B.}\ \bibnamefont {Brookes}}, \bibinfo {author} {\bibfnamefont {G.~M.}\
  \bibnamefont {de~Luca}}, \bibinfo {author} {\bibfnamefont {D.~D.}\
  \bibnamefont {Castro}}, \bibinfo {author} {\bibfnamefont {M.}~\bibnamefont
  {Grioni}}, \bibinfo {author} {\bibfnamefont {M.}~\bibnamefont {Guarise}},
  \bibinfo {author} {\bibfnamefont {P.~G.}\ \bibnamefont {Medaglia}}, \bibinfo
  {author} {\bibfnamefont {F.~M.}\ \bibnamefont {Granozio}}, \bibinfo {author}
  {\bibfnamefont {M.}~\bibnamefont {Minola}}, \bibinfo {author} {\bibfnamefont
  {P.}~\bibnamefont {Perna}}, \bibinfo {author} {\bibfnamefont
  {M.}~\bibnamefont {Radovic}}, \bibinfo {author} {\bibfnamefont
  {M.}~\bibnamefont {Salluzzo}}, \bibinfo {author} {\bibfnamefont
  {T.}~\bibnamefont {Schmitt}}, \bibinfo {author} {\bibfnamefont {K.~J.}\
  \bibnamefont {Zhou}}, \bibinfo {author} {\bibfnamefont {L.}~\bibnamefont
  {Braicovich}}, \ and\ \bibinfo {author} {\bibfnamefont {G.}~\bibnamefont
  {Ghiringhelli}},\ }\href {http://stacks.iop.org/1367-2630/13/i=4/a=043026}
  {\bibfield  {journal} {\bibinfo  {journal} {New Journal of Physics}\ }\textbf
  {\bibinfo {volume} {13}},\ \bibinfo {pages} {043026} (\bibinfo {year}
  {2011})}\BibitemShut {NoStop}%
\bibitem [{\citenamefont {Ament}\ \emph {et~al.}(2011)\citenamefont {Ament},
  \citenamefont {van Veenendaal}, \citenamefont {Devereaux}, \citenamefont
  {Hill},\ and\ \citenamefont {van~den Brink}}]{Ament_RevModPhys.83.705}%
  \BibitemOpen
  \bibfield  {author} {\bibinfo {author} {\bibfnamefont {L.~J.~P.}\
  \bibnamefont {Ament}}, \bibinfo {author} {\bibfnamefont {M.}~\bibnamefont
  {van Veenendaal}}, \bibinfo {author} {\bibfnamefont {T.~P.}\ \bibnamefont
  {Devereaux}}, \bibinfo {author} {\bibfnamefont {J.~P.}\ \bibnamefont {Hill}},
  \ and\ \bibinfo {author} {\bibfnamefont {J.}~\bibnamefont {van~den Brink}},\
  }\href {http://dx.doi.org/10.1103/RevModPhys.83.705} {\bibfield  {journal}
  {\bibinfo  {journal} {Rev. Mod. Phys.}\ }\textbf {\bibinfo {volume} {83}},\
  \bibinfo {pages} {705} (\bibinfo {year} {2011})}\BibitemShut {NoStop}%
\bibitem [{\citenamefont {Braicovich}\ \emph {et~al.}(2010)\citenamefont
  {Braicovich}, \citenamefont {van~den Brink}, \citenamefont {Bisogni},
  \citenamefont {Sala}, \citenamefont {Ament}, \citenamefont {Brookes},
  \citenamefont {De~Luca}, \citenamefont {Salluzzo}, \citenamefont {Schmitt},
  \citenamefont {Strocov},\ and\ \citenamefont
  {Ghiringhelli}}]{Lucio_PhysRevLett.104.077002}%
  \BibitemOpen
  \bibfield  {author} {\bibinfo {author} {\bibfnamefont {L.}~\bibnamefont
  {Braicovich}}, \bibinfo {author} {\bibfnamefont {J.}~\bibnamefont {van~den
  Brink}}, \bibinfo {author} {\bibfnamefont {V.}~\bibnamefont {Bisogni}},
  \bibinfo {author} {\bibfnamefont {M.~M.}\ \bibnamefont {Sala}}, \bibinfo
  {author} {\bibfnamefont {L.~J.~P.}\ \bibnamefont {Ament}}, \bibinfo {author}
  {\bibfnamefont {N.~B.}\ \bibnamefont {Brookes}}, \bibinfo {author}
  {\bibfnamefont {G.~M.}\ \bibnamefont {De~Luca}}, \bibinfo {author}
  {\bibfnamefont {M.}~\bibnamefont {Salluzzo}}, \bibinfo {author}
  {\bibfnamefont {T.}~\bibnamefont {Schmitt}}, \bibinfo {author} {\bibfnamefont
  {V.~N.}\ \bibnamefont {Strocov}}, \ and\ \bibinfo {author} {\bibfnamefont
  {G.}~\bibnamefont {Ghiringhelli}},\ }\href
  {http://dx.doi.org/10.1103/PhysRevLett.104.077002} {\bibfield  {journal}
  {\bibinfo  {journal} {Phys. Rev. Lett.}\ }\textbf {\bibinfo {volume} {104}},\
  \bibinfo {pages} {077002} (\bibinfo {year} {2010})}\BibitemShut {NoStop}%
\bibitem [{\citenamefont {Le~Tacon}\ \emph {et~al.}(2011)\citenamefont
  {Le~Tacon}, \citenamefont {Ghiringhelli}, \citenamefont {Chaloupka},
  \citenamefont {Sala}, \citenamefont {Hinkov}, \citenamefont {Haverkort},
  \citenamefont {Minola}, \citenamefont {Bakr}, \citenamefont {Zhou},
  \citenamefont {Blanco-Canosa}, \citenamefont {Monney}, \citenamefont {Song},
  \citenamefont {Sun}, \citenamefont {Lin}, \citenamefont {De~Luca},
  \citenamefont {Salluzzo}, \citenamefont {Khaliullin}, \citenamefont
  {Schmitt}, \citenamefont {Braicovich},\ and\ \citenamefont
  {Keimer}}]{LeTacon2011}%
  \BibitemOpen
  \bibfield  {author} {\bibinfo {author} {\bibfnamefont {M.}~\bibnamefont
  {Le~Tacon}}, \bibinfo {author} {\bibfnamefont {G.}~\bibnamefont
  {Ghiringhelli}}, \bibinfo {author} {\bibfnamefont {J.}~\bibnamefont
  {Chaloupka}}, \bibinfo {author} {\bibfnamefont {M.~M.}\ \bibnamefont {Sala}},
  \bibinfo {author} {\bibfnamefont {V.}~\bibnamefont {Hinkov}}, \bibinfo
  {author} {\bibfnamefont {M.~W.}\ \bibnamefont {Haverkort}}, \bibinfo {author}
  {\bibfnamefont {M.}~\bibnamefont {Minola}}, \bibinfo {author} {\bibfnamefont
  {M.}~\bibnamefont {Bakr}}, \bibinfo {author} {\bibfnamefont {K.~J.}\
  \bibnamefont {Zhou}}, \bibinfo {author} {\bibfnamefont {S.}~\bibnamefont
  {Blanco-Canosa}}, \bibinfo {author} {\bibfnamefont {C.}~\bibnamefont
  {Monney}}, \bibinfo {author} {\bibfnamefont {Y.~T.}\ \bibnamefont {Song}},
  \bibinfo {author} {\bibfnamefont {G.~L.}\ \bibnamefont {Sun}}, \bibinfo
  {author} {\bibfnamefont {C.~T.}\ \bibnamefont {Lin}}, \bibinfo {author}
  {\bibfnamefont {G.~M.}\ \bibnamefont {De~Luca}}, \bibinfo {author}
  {\bibfnamefont {M.}~\bibnamefont {Salluzzo}}, \bibinfo {author}
  {\bibfnamefont {G.}~\bibnamefont {Khaliullin}}, \bibinfo {author}
  {\bibfnamefont {T.}~\bibnamefont {Schmitt}}, \bibinfo {author} {\bibfnamefont
  {L.}~\bibnamefont {Braicovich}}, \ and\ \bibinfo {author} {\bibfnamefont
  {B.}~\bibnamefont {Keimer}},\ }\href {http://dx.doi.org/10.1038/nphys2041}
  {\bibfield  {journal} {\bibinfo  {journal} {Nat. Phys.}\ }\textbf {\bibinfo
  {volume} {7}},\ \bibinfo {pages} {725} (\bibinfo {year} {2011})}\BibitemShut
  {NoStop}%
\bibitem [{\citenamefont {Lee}\ \emph {et~al.}(2014)\citenamefont {Lee},
  \citenamefont {Lee}, \citenamefont {Nowadnick}, \citenamefont {Gerber},
  \citenamefont {Tabis}, \citenamefont {Huang}, \citenamefont {Strocov},
  \citenamefont {Motoyama}, \citenamefont {Yu}, \citenamefont {Moritz},
  \citenamefont {Huang}, \citenamefont {Wang}, \citenamefont {Huang},
  \citenamefont {Wu}, \citenamefont {Chen}, \citenamefont {Huang},
  \citenamefont {Greven}, \citenamefont {Schmitt}, \citenamefont {Shen},\ and\
  \citenamefont {Devereaux}}]{Lee2014}%
  \BibitemOpen
  \bibfield  {author} {\bibinfo {author} {\bibfnamefont {W.~S.}\ \bibnamefont
  {Lee}}, \bibinfo {author} {\bibfnamefont {J.~J.}\ \bibnamefont {Lee}},
  \bibinfo {author} {\bibfnamefont {E.~A.}\ \bibnamefont {Nowadnick}}, \bibinfo
  {author} {\bibfnamefont {S.}~\bibnamefont {Gerber}}, \bibinfo {author}
  {\bibfnamefont {W.}~\bibnamefont {Tabis}}, \bibinfo {author} {\bibfnamefont
  {S.~W.}\ \bibnamefont {Huang}}, \bibinfo {author} {\bibfnamefont {V.~N.}\
  \bibnamefont {Strocov}}, \bibinfo {author} {\bibfnamefont {E.~M.}\
  \bibnamefont {Motoyama}}, \bibinfo {author} {\bibfnamefont {G.}~\bibnamefont
  {Yu}}, \bibinfo {author} {\bibfnamefont {B.}~\bibnamefont {Moritz}}, \bibinfo
  {author} {\bibfnamefont {H.~Y.}\ \bibnamefont {Huang}}, \bibinfo {author}
  {\bibfnamefont {R.~P.}\ \bibnamefont {Wang}}, \bibinfo {author}
  {\bibfnamefont {Y.~B.}\ \bibnamefont {Huang}}, \bibinfo {author}
  {\bibfnamefont {W.~B.}\ \bibnamefont {Wu}}, \bibinfo {author} {\bibfnamefont
  {C.~T.}\ \bibnamefont {Chen}}, \bibinfo {author} {\bibfnamefont {D.~J.}\
  \bibnamefont {Huang}}, \bibinfo {author} {\bibfnamefont {M.}~\bibnamefont
  {Greven}}, \bibinfo {author} {\bibfnamefont {T.}~\bibnamefont {Schmitt}},
  \bibinfo {author} {\bibfnamefont {Z.~X.}\ \bibnamefont {Shen}}, \ and\
  \bibinfo {author} {\bibfnamefont {T.~P.}\ \bibnamefont {Devereaux}},\ }\href
  {http://dx.doi.org/10.1038/nphys3117} {\bibfield  {journal} {\bibinfo
  {journal} {Nat Phys}\ }\textbf {\bibinfo {volume} {10}},\ \bibinfo {pages}
  {883} (\bibinfo {year} {2014})}\BibitemShut {NoStop}%
\bibitem [{\citenamefont {Peng}\ \emph {et~al.}(2015)\citenamefont {Peng},
  \citenamefont {Hashimoto}, \citenamefont {Sala}, \citenamefont {Amorese},
  \citenamefont {Brookes}, \citenamefont {Dellea}, \citenamefont {Lee},
  \citenamefont {Minola}, \citenamefont {Schmitt}, \citenamefont {Yoshida},
  \citenamefont {Zhou}, \citenamefont {Eisaki}, \citenamefont {Devereaux},
  \citenamefont {Shen}, \citenamefont {Braicovich},\ and\ \citenamefont
  {Ghiringhelli}}]{YY_Peng_2015}%
  \BibitemOpen
  \bibfield  {author} {\bibinfo {author} {\bibfnamefont {Y.~Y.}\ \bibnamefont
  {Peng}}, \bibinfo {author} {\bibfnamefont {M.}~\bibnamefont {Hashimoto}},
  \bibinfo {author} {\bibfnamefont {M.~M.}\ \bibnamefont {Sala}}, \bibinfo
  {author} {\bibfnamefont {A.}~\bibnamefont {Amorese}}, \bibinfo {author}
  {\bibfnamefont {N.~B.}\ \bibnamefont {Brookes}}, \bibinfo {author}
  {\bibfnamefont {G.}~\bibnamefont {Dellea}}, \bibinfo {author} {\bibfnamefont
  {W.-S.}\ \bibnamefont {Lee}}, \bibinfo {author} {\bibfnamefont
  {M.}~\bibnamefont {Minola}}, \bibinfo {author} {\bibfnamefont
  {T.}~\bibnamefont {Schmitt}}, \bibinfo {author} {\bibfnamefont
  {Y.}~\bibnamefont {Yoshida}}, \bibinfo {author} {\bibfnamefont {K.-J.}\
  \bibnamefont {Zhou}}, \bibinfo {author} {\bibfnamefont {H.}~\bibnamefont
  {Eisaki}}, \bibinfo {author} {\bibfnamefont {T.~P.}\ \bibnamefont
  {Devereaux}}, \bibinfo {author} {\bibfnamefont {Z.-X.}\ \bibnamefont {Shen}},
  \bibinfo {author} {\bibfnamefont {L.}~\bibnamefont {Braicovich}}, \ and\
  \bibinfo {author} {\bibfnamefont {G.}~\bibnamefont {Ghiringhelli}},\ }\href
  {\doibase 10.1103/PhysRevB.92.064517} {\bibfield  {journal} {\bibinfo
  {journal} {Phys. Rev. B}\ }\textbf {\bibinfo {volume} {92}},\ \bibinfo
  {pages} {064517} (\bibinfo {year} {2015})}\BibitemShut {NoStop}%
\bibitem [{\citenamefont {Jia}\ \emph {et~al.}(2016)\citenamefont {Jia},
  \citenamefont {Wohlfeld}, \citenamefont {Wang}, \citenamefont {Moritz},\ and\
  \citenamefont {Devereaux}}]{Jia_PhysRevX.6.021020}%
  \BibitemOpen
  \bibfield  {author} {\bibinfo {author} {\bibfnamefont {C.}~\bibnamefont
  {Jia}}, \bibinfo {author} {\bibfnamefont {K.}~\bibnamefont {Wohlfeld}},
  \bibinfo {author} {\bibfnamefont {Y.}~\bibnamefont {Wang}}, \bibinfo {author}
  {\bibfnamefont {B.}~\bibnamefont {Moritz}}, \ and\ \bibinfo {author}
  {\bibfnamefont {T.~P.}\ \bibnamefont {Devereaux}},\ }\href
  {http://dx.doi.org/10.1103/PhysRevX.6.021020} {\bibfield  {journal} {\bibinfo
   {journal} {Phys. Rev. X}\ }\textbf {\bibinfo {volume} {6}},\ \bibinfo
  {pages} {021020} (\bibinfo {year} {2016})}\BibitemShut {NoStop}%
\bibitem [{\citenamefont {Devereaux}\ \emph {et~al.}(2016)\citenamefont
  {Devereaux}, \citenamefont {Shvaika}, \citenamefont {Wu}, \citenamefont
  {Wohlfeld}, \citenamefont {Jia}, \citenamefont {Wang}, \citenamefont
  {Moritz}, \citenamefont {Chaix}, \citenamefont {Lee}, \citenamefont {Shen},
  \citenamefont {Ghiringhelli},\ and\ \citenamefont
  {Braicovich}}]{Devereaux_PhysRevX.6.041019}%
  \BibitemOpen
  \bibfield  {author} {\bibinfo {author} {\bibfnamefont {T.~P.}\ \bibnamefont
  {Devereaux}}, \bibinfo {author} {\bibfnamefont {A.~M.}\ \bibnamefont
  {Shvaika}}, \bibinfo {author} {\bibfnamefont {K.}~\bibnamefont {Wu}},
  \bibinfo {author} {\bibfnamefont {K.}~\bibnamefont {Wohlfeld}}, \bibinfo
  {author} {\bibfnamefont {C.~J.}\ \bibnamefont {Jia}}, \bibinfo {author}
  {\bibfnamefont {Y.}~\bibnamefont {Wang}}, \bibinfo {author} {\bibfnamefont
  {B.}~\bibnamefont {Moritz}}, \bibinfo {author} {\bibfnamefont
  {L.}~\bibnamefont {Chaix}}, \bibinfo {author} {\bibfnamefont {W.-S.}\
  \bibnamefont {Lee}}, \bibinfo {author} {\bibfnamefont {Z.-X.}\ \bibnamefont
  {Shen}}, \bibinfo {author} {\bibfnamefont {G.}~\bibnamefont {Ghiringhelli}},
  \ and\ \bibinfo {author} {\bibfnamefont {L.}~\bibnamefont {Braicovich}},\
  }\href {\doibase 10.1103/PhysRevX.6.041019} {\bibfield  {journal} {\bibinfo
  {journal} {Phys. Rev. X}\ }\textbf {\bibinfo {volume} {6}},\ \bibinfo {pages}
  {041019} (\bibinfo {year} {2016})}\BibitemShut {NoStop}%
\bibitem [{\citenamefont {Hill}\ \emph {et~al.}(2008)\citenamefont {Hill},
  \citenamefont {Blumberg}, \citenamefont {Kim}, \citenamefont {Ellis},
  \citenamefont {Wakimoto}, \citenamefont {Birgeneau}, \citenamefont {Komiya},
  \citenamefont {Ando}, \citenamefont {Liang}, \citenamefont {Greene},
  \citenamefont {Casa},\ and\ \citenamefont
  {Gog}}]{Hill_PhysRevLett.100.097001}%
  \BibitemOpen
  \bibfield  {author} {\bibinfo {author} {\bibfnamefont {J.~P.}\ \bibnamefont
  {Hill}}, \bibinfo {author} {\bibfnamefont {G.}~\bibnamefont {Blumberg}},
  \bibinfo {author} {\bibfnamefont {Y.-J.}\ \bibnamefont {Kim}}, \bibinfo
  {author} {\bibfnamefont {D.~S.}\ \bibnamefont {Ellis}}, \bibinfo {author}
  {\bibfnamefont {S.}~\bibnamefont {Wakimoto}}, \bibinfo {author}
  {\bibfnamefont {R.~J.}\ \bibnamefont {Birgeneau}}, \bibinfo {author}
  {\bibfnamefont {S.}~\bibnamefont {Komiya}}, \bibinfo {author} {\bibfnamefont
  {Y.}~\bibnamefont {Ando}}, \bibinfo {author} {\bibfnamefont {B.}~\bibnamefont
  {Liang}}, \bibinfo {author} {\bibfnamefont {R.~L.}\ \bibnamefont {Greene}},
  \bibinfo {author} {\bibfnamefont {D.}~\bibnamefont {Casa}}, \ and\ \bibinfo
  {author} {\bibfnamefont {T.}~\bibnamefont {Gog}},\ }\href {\doibase
  10.1103/PhysRevLett.100.097001} {\bibfield  {journal} {\bibinfo  {journal}
  {Phys. Rev. Lett.}\ }\textbf {\bibinfo {volume} {100}},\ \bibinfo {pages}
  {097001} (\bibinfo {year} {2008})}\BibitemShut {NoStop}%
\bibitem [{\citenamefont {{Ishii}}\ \emph {et~al.}(2014)\citenamefont
  {{Ishii}}, \citenamefont {{Fujita}}, \citenamefont {{Sasaki}}, \citenamefont
  {{Minola}}, \citenamefont {{Dellea}}, \citenamefont {{Mazzoli}},
  \citenamefont {{Kummer}}, \citenamefont {{Ghiringhelli}}, \citenamefont
  {{Braicovich}}, \citenamefont {{Tohyama}}, \citenamefont {{Tsutsumi}},
  \citenamefont {{Sato}}, \citenamefont {{Kajimoto}}, \citenamefont
  {{Ikeuchi}}, \citenamefont {{Yamada}}, \citenamefont {{Yoshida}},
  \citenamefont {{Kurooka}},\ and\ \citenamefont {{Mizuki}}}]{Ishii_Soin_2014}%
  \BibitemOpen
  \bibfield  {author} {\bibinfo {author} {\bibfnamefont {K.}~\bibnamefont
  {{Ishii}}}, \bibinfo {author} {\bibfnamefont {M.}~\bibnamefont {{Fujita}}},
  \bibinfo {author} {\bibfnamefont {T.}~\bibnamefont {{Sasaki}}}, \bibinfo
  {author} {\bibfnamefont {M.}~\bibnamefont {{Minola}}}, \bibinfo {author}
  {\bibfnamefont {G.}~\bibnamefont {{Dellea}}}, \bibinfo {author}
  {\bibfnamefont {C.}~\bibnamefont {{Mazzoli}}}, \bibinfo {author}
  {\bibfnamefont {K.}~\bibnamefont {{Kummer}}}, \bibinfo {author}
  {\bibfnamefont {G.}~\bibnamefont {{Ghiringhelli}}}, \bibinfo {author}
  {\bibfnamefont {L.}~\bibnamefont {{Braicovich}}}, \bibinfo {author}
  {\bibfnamefont {T.}~\bibnamefont {{Tohyama}}}, \bibinfo {author}
  {\bibfnamefont {K.}~\bibnamefont {{Tsutsumi}}}, \bibinfo {author}
  {\bibfnamefont {K.}~\bibnamefont {{Sato}}}, \bibinfo {author} {\bibfnamefont
  {R.}~\bibnamefont {{Kajimoto}}}, \bibinfo {author} {\bibfnamefont
  {K.}~\bibnamefont {{Ikeuchi}}}, \bibinfo {author} {\bibfnamefont
  {K.}~\bibnamefont {{Yamada}}}, \bibinfo {author} {\bibfnamefont
  {M.}~\bibnamefont {{Yoshida}}}, \bibinfo {author} {\bibfnamefont
  {M.}~\bibnamefont {{Kurooka}}}, \ and\ \bibinfo {author} {\bibfnamefont
  {J.}~\bibnamefont {{Mizuki}}},\ }\href@noop {} {\bibfield  {journal}
  {\bibinfo  {journal} {Nature Communications}\ }\textbf {\bibinfo {volume}
  {5}},\ \bibinfo {eid} {3714} (\bibinfo {year} {2014})}\BibitemShut {NoStop}%
\bibitem [{\citenamefont {d'Astuto}\ \emph {et~al.}(2002)\citenamefont
  {d'Astuto}, \citenamefont {Mang}, \citenamefont {Giura}, \citenamefont
  {Shukla}, \citenamefont {Ghigna}, \citenamefont {Mirone}, \citenamefont
  {Braden}, \citenamefont {Greven}, \citenamefont {Krisch},\ and\ \citenamefont
  {Sette}}]{dAstuto_2002}%
  \BibitemOpen
  \bibfield  {author} {\bibinfo {author} {\bibfnamefont {M.}~\bibnamefont
  {d'Astuto}}, \bibinfo {author} {\bibfnamefont {P.~K.}\ \bibnamefont {Mang}},
  \bibinfo {author} {\bibfnamefont {P.}~\bibnamefont {Giura}}, \bibinfo
  {author} {\bibfnamefont {A.}~\bibnamefont {Shukla}}, \bibinfo {author}
  {\bibfnamefont {P.}~\bibnamefont {Ghigna}}, \bibinfo {author} {\bibfnamefont
  {A.}~\bibnamefont {Mirone}}, \bibinfo {author} {\bibfnamefont
  {M.}~\bibnamefont {Braden}}, \bibinfo {author} {\bibfnamefont
  {M.}~\bibnamefont {Greven}}, \bibinfo {author} {\bibfnamefont
  {M.}~\bibnamefont {Krisch}}, \ and\ \bibinfo {author} {\bibfnamefont
  {F.}~\bibnamefont {Sette}},\ }\href {\doibase 10.1103/PhysRevLett.88.167002}
  {\bibfield  {journal} {\bibinfo  {journal} {Phys. Rev. Lett.}\ }\textbf
  {\bibinfo {volume} {88}},\ \bibinfo {pages} {167002} (\bibinfo {year}
  {2002})}\BibitemShut {NoStop}%
\bibitem [{\citenamefont {Bejas}\ \emph {et~al.}(2017)\citenamefont {Bejas},
  \citenamefont {Yamase},\ and\ \citenamefont {Greco}}]{Greco_PRB}%
  \BibitemOpen
  \bibfield  {author} {\bibinfo {author} {\bibfnamefont {M.}~\bibnamefont
  {Bejas}}, \bibinfo {author} {\bibfnamefont {H.}~\bibnamefont {Yamase}}, \
  and\ \bibinfo {author} {\bibfnamefont {A.}~\bibnamefont {Greco}},\ }\href
  {\doibase 10.1103/PhysRevB.96.214513} {\bibfield  {journal} {\bibinfo
  {journal} {Phys. Rev. B}\ }\textbf {\bibinfo {volume} {96}},\ \bibinfo
  {pages} {214513} (\bibinfo {year} {2017})}\BibitemShut {NoStop}%
\bibitem [{\citenamefont {Braicovich}\ \emph {et~al.}(2014)\citenamefont
  {Braicovich}, \citenamefont {Minola}, \citenamefont {Dellea}, \citenamefont
  {Tacon}, \citenamefont {Sala}, \citenamefont {Morawe}, \citenamefont
  {Peffen}, \citenamefont {Supruangnet}, \citenamefont {Yakhou}, \citenamefont
  {Ghiringhelli},\ and\ \citenamefont {Brookes}}]{LB_RSI}%
  \BibitemOpen
  \bibfield  {author} {\bibinfo {author} {\bibfnamefont {L.}~\bibnamefont
  {Braicovich}}, \bibinfo {author} {\bibfnamefont {M.}~\bibnamefont {Minola}},
  \bibinfo {author} {\bibfnamefont {G.}~\bibnamefont {Dellea}}, \bibinfo
  {author} {\bibfnamefont {M.~L.}\ \bibnamefont {Tacon}}, \bibinfo {author}
  {\bibfnamefont {M.~M.}\ \bibnamefont {Sala}}, \bibinfo {author}
  {\bibfnamefont {C.}~\bibnamefont {Morawe}}, \bibinfo {author} {\bibfnamefont
  {J.-C.}\ \bibnamefont {Peffen}}, \bibinfo {author} {\bibfnamefont
  {R.}~\bibnamefont {Supruangnet}}, \bibinfo {author} {\bibfnamefont
  {F.}~\bibnamefont {Yakhou}}, \bibinfo {author} {\bibfnamefont
  {G.}~\bibnamefont {Ghiringhelli}}, \ and\ \bibinfo {author} {\bibfnamefont
  {N.~B.}\ \bibnamefont {Brookes}},\ }\href {\doibase 10.1063/1.4900959}
  {\bibfield  {journal} {\bibinfo  {journal} {Review of Scientific
  Instruments}\ }\textbf {\bibinfo {volume} {85}},\ \bibinfo {pages} {115104}
  (\bibinfo {year} {2014})}\BibitemShut {NoStop}%
\bibitem [{\citenamefont {Li}\ \emph {et~al.}(2017)\citenamefont {Li},
  \citenamefont {Wang}, \citenamefont {Yao},\ and\ \citenamefont
  {Lee}}]{Li_2017}%
  \BibitemOpen
  \bibfield  {author} {\bibinfo {author} {\bibfnamefont {Z.-X.}\ \bibnamefont
  {Li}}, \bibinfo {author} {\bibfnamefont {F.}~\bibnamefont {Wang}}, \bibinfo
  {author} {\bibfnamefont {H.}~\bibnamefont {Yao}}, \ and\ \bibinfo {author}
  {\bibfnamefont {D.-H.}\ \bibnamefont {Lee}},\ }\href {\doibase
  10.1103/PhysRevB.95.214505} {\bibfield  {journal} {\bibinfo  {journal} {Phys.
  Rev. B}\ }\textbf {\bibinfo {volume} {95}},\ \bibinfo {pages} {214505}
  (\bibinfo {year} {2017})}\BibitemShut {NoStop}%
\end{thebibliography}%
\end{document}